\documentclass[iop,apj,twocolumn,twocolappendix,numberedappendix]{emulateapj}
\usepackage{latexsym,graphicx,rotating,amsmath, epsfig, natbib}
\usepackage[usenames, dvipsnames]{color}

\newcommand\Beq{\begin{eqnarray}} 
\newcommand\Eeq{\end{eqnarray}}

\newcommand{\n}{  \\ }
\newcommand{\nn}{\nonumber}

\newcommand{\eq}[1]{equation~(\ref{#1})}
\newcommand{\eqs}[2]{equations~(\ref{#1})~\&~(\ref{#2})}
\newcommand{\eqss}[2]{equations~(\ref{#1})--(\ref{#2})}
\newcommand{\Eq}[1]{Equation~(\ref{#1})}
\newcommand{\Eqs}[2]{Equations~(\ref{#1})~\&~(\ref{#2})}
\newcommand{\Eqss}[2]{Equations~(\ref{#1})--(\ref{#2})}
\renewcommand{\sec}[1]{\S\ref{sec: #1}}
\newcommand{\fig}[2]{Figure~\ref{#1}{\textit{#2}}}

\newcommand{\ie}{i.e., }

\renewcommand{\O}[1]{\mathcal{O}{(#1)}}
\newcommand{\C}{\mathcal{C}}
\newcommand{\Ma}{\mbox{\textit{Ma}}}
\newcommand{\PE}{\mbox{\textit{PE}}}
\newcommand{\PK}{\mbox{\textit{PK}}}
\newcommand{\PU}{\mbox{\textit{PU}}}

\newcommand{\ann}{\citep{Almgren_2000, 
Almgren_et_al_2006a_ApJ, Almgren_et_al_2006b_ApJ, 
Almgren_et_al_2008}}

\renewcommand{\vec}[1]{\boldsymbol{#1}}
\renewcommand{\u}{\vec{u}}
\newcommand{\g}{\vec{g}}
\newcommand{\vxi}{\vec{\xi}}

\newcommand{\dd}[1]{\,\mathrm{d}{#1}}
\newcommand{\dV}{\,\mathrm{d}^{3}\vec{x}}
\newcommand{\dt}{\,\mathrm{d}t }

\newcommand{\inv}[1]{\frac{1}{#1}}
\renewcommand{\dot}{\vec{\cdot}}
\newcommand{\cross}{\vec{\times}}

\newcommand{\del}{\nabla}
\newcommand{\grad}{\vec{\nabla}}
\renewcommand{\div}{ \grad \dot }
\newcommand{\curl}{ \grad \cross }

\renewcommand{\perp}{\!\bot}
\newcommand{\delsq}{\del^2_{\perp}}

\newcommand{\pd}[2]{\frac{\partial{#1}}{\partial {#2}}}
\newcommand{\thermod}[3]{\bigg(\pd{#1}{#2}\bigg)_{\!\! #3}\!}
\newcommand{\Dt}[1]{\pd{#1}{t} + \u \dot \grad {#1}}

\newcommand\JFM{\rmfamily{J. Fluid Mech.}}

\shorttitle{Lagrangian Constrained Flow}
\shortauthors{Vasil, Lecoanet, Brown, Wood \& Zweibel} 

\begin{document}

\title{Energy Conservation and Gravity Waves in Sound-proof Treatments of Stellar Interiors: Part II Lagrangian Constrained Analysis \bigskip}
\author{Geoffrey M.\ Vasil}
\affil{Dept.\ Astronomy \& Theoretical Astrophysics Center, University of California Berkeley, Berkeley, CA 94720}
\affil{Canadian Institute for Theoretical Astrophysics, University of Toronto, 60 St.~George Street, Toronto, ON M5S~3H8 Canada}
\email{vasil@cita.utoronto.ca}
\author{Daniel Lecoanet}
\affil{Dept.\ Astronomy \& Theoretical Astrophysics Center, University of California Berkeley, Berkeley, CA 94720}
\author{Benjamin P.\ Brown}
\affil{Dept.\ Astronomy, University of Wisconsin, Madison, WI 53706-1582}
\affil{Center for Magnetic Self Organization in Laboratory and Astrophysical Plasmas, University of Wisconsin, 1150 University Avenue, Madison, WI 53706, USA}
\author{Toby S.\ Wood}
\affil{Department of Applied Mathematics and Statistics, Baskin School of Engineering, University of California Santa Cruz, CA, USA}
\author{Ellen G.\ Zweibel}
\affil{Dept.\ Astronomy, University of Wisconsin, Madison, WI 53706-1582}
\affil{Center for Magnetic Self Organization in Laboratory and Astrophysical Plasmas, University of Wisconsin, 1150 University Avenue, Madison, WI 53706, USA}

\begin{abstract}
The speed of sound greatly exceeds typical flow velocities in many stellar and planetary interiors.  To follow the slow evolution of subsonic motions, various sound-proof models attempt to remove fast acoustic waves whilst retaining stratified convection and buoyancy dynamics.  In astrophysics, anelastic models typically receive the most attention in the class of sound-filtered stratified models.  Generally, anelastic models remain valid in nearly adiabatically stratified regions like stellar convection zones, but may break down in strongly sub-adiabatic, stably stratified layers common in stellar radiative zones.  However, studying stellar rotation, circulation, and dynamos requires understanding the complex coupling between convection and radiative zones, and this requires robust equations valid in both regimes. Here we extend the analysis of equation sets begun in \citet{Brown_Vasil_Zweibel_2012}, which studied anelastic models, to two types of pseudo-incompressible models.  This class of models has received attention in atmospheric applications, and more recently in studies of white-dwarf supernovae progenitors.  We demonstrate that one model conserves energy but the other does not. We use Lagrangian variational methods to extend the energy conserving model to a general equation of state, and dub the resulting equation set the Generalized Pseudo-Incompressible (GPI) model. We show that the GPI equations suitably capture low frequency phenomena in both convection and radiative zones in stars and other stratified systems, and we provide recommendations for converting low-Mach number codes to this equation set.
 \end{abstract}
\slugcomment{}


\section{Introduction and motivation}\label{sec: Introduction and motivation}

In astrophysical fluid dynamics, the relevant  timescales are often substantially longer than the sound crossing time of the system.  This particularly holds true for convection deep in stellar interiors where the flows are very subsonic.  Near the base of the solar convection zone the sound speed is about 220 km/s, while the convective velocities are likely of order hundreds of meters per
second.  Following the evolution of sound directly imposes crippling computational limits on simulations of such flows, as their evolution
times are typically many convective turnover times, each of which is often several thousand sound times.

For numerical stability, an explicit time-integration scheme must satisfy the Courant--Friedrichs--Lewy (CFL) condition, that the time-step must be smaller than the shortest timescale admitted by the equations. 
For the non-rotating Euler equations in a stably stratified medium, this is
\Beq
\label{CFL}
\Delta t \lesssim \min\left\{{\frac{\Delta x}{c_{0}}, \frac{\Delta x}{u_{0}}, \inv{N_{0}}}\right\},
\Eeq
where $\Delta t$ represents the time-step size, $\Delta x$ is the smallest resolved length scale in the calculation, and $c_0$, $u_0$, and $N_0$ represent characteristic sound speed, flow velocity, and Brunt-V\"ais\"al\"a frequency respectively.  The Mach number gives the ratio of the first two timescales in \eq{CFL}
\Beq
\Ma \equiv \frac{|\u_{0}|}{c_{0}},
\Eeq
which is typically very small in stellar interiors.  If we
make the order of magnitude estimates $c_0^{2} \sim g H $ and $N_0^{2} \sim g/H $, where $g$, and $H$ represent gravitational acceleration and density scale height respectively, then
the ratio of the first and third timescales in \eq{CFL} is
\Beq
\frac{\Delta x}{H},
\Eeq
which is also small in high-resolution simulations of stellar interiors.

So called ``sound-proof'' models address this separation of time scales by starting with the Euler equations and filtering out fast,
high-frequency sound waves while retaining compressible motions on slower time scales due to gravitational stratification.  These motions
include gravity waves in stably stratified regions and convection in unstably stratified regions.  The CFL condition for sound-proof equations only requires
\Beq
\Delta t \lesssim \min\left\{\frac{\Delta x}{u_{0}}, \inv{N_{0}}\right\},
\Eeq
which much-less severely restricts efficiency than \eq{CFL} for the fully compressible Euler equations.  In astrophysical and geophysical settings,  the most commonly employed ``sound-proof'' models are the anelastic equations \citep{Batchelor_1953, Ogura&Phillips_1962, Gough_1969}. Fundamentally, anelastic models filter sound waves by modifying the continuity equation of the fully compressible Euler equations.  Formally, this anelastic approximation is only valid for an adiabatic or nearly adiabatic atmosphere \citep{Gough_1969,Braginsky&Roberts_1995}.

The pseudo-incompressible models --- with a modified pressure, rather than density, equation --- give an alternate approach to sound proofing the Navier--Stokes equations.  
\citet{Durran_1989} first proposed this class of models, which the astrophysical fluid dynamics community more recently adapted to a number of applications \citep[e.g.,][]{Almgren_et_al_2006a_ApJ, 
Almgren_et_al_2006b_ApJ, Zingale_et_al_2009}.  A pseudo-incompressible model finds particular use in the MAESTRO code \citep{Nonaka_et_al_2010}.  
The atmospheric sciences community has also extensively explored the properties of gravity waves and stable-layer dynamics in
pseudo-incompressible models, with several comparisons against the
properties of anelastic models
\citep{Durran_1989, Durran_2008, Nance&Durran_1994, Achatz_et_al_2010,
Klein_et_al_2010}.  

In \citet{Brown_Vasil_Zweibel_2012} (hereinafter Part~I) we discussed the
energy conserving properties of various anelastic models in stably
stratified, sub-adiabatic atmospheres such as are found in stellar
radiative zones.  We found that some widely used anelastic models violate energy conservation.  This behaviour escaped attention previously as in bounded atmospheres, the norm for
most simulations, energy-violating anelastic models instead
conserve a pseudo-energy (an energy-like quadratic invariant with incorrect stratification weighting).  Internal gravity wave eigenfunctions in
energy-violating anelastic models can differ significantly from the fully compressible results, and neither energy nor pseudo-energy remain conserved when nonlinear dynamics become important. Anelastic models that correctly conserve energy have modified momentum equations, a theme which reoccurs here in our study of pseudo-incompressible models.

Here we explore two pseudo-incompressible
models (\sec{Background and model equations}), one used in the atmospheric community (PI equations),
and a new variant being used in the astrophysical community that includes more general equations of state (LM equations).
We find that the PI equations conserve energy but that the LM equations do not.
We additionally derive a new subsonic flow model based on a
constrained Lagrangian analysis of the full compressible flow
(\sec{Lagrangian Analysis} \& \ref{sec: Constrained Models}).  
We call these the Generalized Pseudo-Incompressible (GPI) equations.
The GPI equations conserves energy and 
correctly generalizes the pseudo-incompressible approximation
for an arbitrary equation of state including astrophysically relevant
situations such as radiation hydrodynamics. 
After analyzing the general properties of the GPI, PI,  and LM equations, 
we specialize to the case of an ideal gas equation of state to explore the behavior of these differing models in bounded
atmospheres (\sec{isothermal atmosphere}) and perform numerical simulations that show the difference
between the LM and GPI equations (\sec{bounded atmospheres and numerics}).
The implications of these findings for simulations of stellar interiors is discussed in
\sec{conclusions}.
The essential results of this paper are the derivation of the
Euler--Lagrange equation~(\ref{Euler--Lagrange}) for any general
non-dissipative fluid flow, and application of
equation~(\ref{Euler--Lagrange})  to derive the 
Generalized Psuedo-Incompressible equations as expressed 
in \eqss{eq:new-momentum}{eq:new-pressure}.
The reader who is primarily interested in implementing
energy-conserving subsonic models with a general equation of state 
should read \sec{Lagrangian Analysis}, \ref{sec: Constrained Models}, 
\ref{sec: bounded atmospheres and numerics}
and \ref{sec: conclusions}.

\section{Background and model equations}
\label{sec: Background and model equations}

\subsection{Thermodynamics and Stratification}
\label{sec: Thermodynamics and Stratification}

We begin our discussion with thermodynamics 
and the geometric conservation laws of mass and entropy 
\Beq
\label{rho-eq} &&\pd{\rho}{t} + \div{\left(\rho \u\right)} \ = \ 0\n
\label{s-eq} &&\Dt{s} \ = \ 0,
\Eeq
where $\u (t,\vec{x})$ denotes the Eulerian (fixed spatial coordinates) fluid velocity, $\rho(t,\vec{x})$ denotes the density, and $s(t,\vec{x})$ represents the specific entropy.  The main result of this paper provides a framework for producing appropriate evolution equations for the fluid velocity under different physical assumptions and approximations.  At this stage, $\u (t,\vec{x})$ represents an arbitrary flow field.  
 
The combined First and Second Laws of Thermodynamics relate small changes in density, entropy and specific internal energy, $e$, via, 
\Beq
\label{2nd Law} T \dd{s} = \dd{e} - \frac{p}{\rho^{2}} \dd{\rho},
\Eeq
\Eq{2nd Law} suggests that the internal energy  naturally depends on the density and entropy.  An equation of state closes the thermodynamic description of the system, which comes in the form of an algebraic (instantaneously valid at every point in space) relation between any three thermodynamic variables; for example $p = p(\rho,s)$, or more fundamentally $e = e(\rho,s)$.  Therefore, 
\Beq
\label{p(rho,s)-eq} p(\rho,s) \equiv  \rho^{2} \thermod{e}{\rho}{s},\quad T(\rho,s) \equiv  \thermod{e}{s}{\rho}.
\Eeq 
define the pressure, $p$, and temperature, $T$. We consider the two partial derivatives of internal energy $e$ at constant entropy $s$ and density $\rho$ respectively.   

We also find it useful to define the specific enthalpy,
\Beq
\label{Enthalpy 2nd Law}h \equiv e + \frac{p}{\rho}, \quad \dd{h} =   T \dd{s} +  \frac{\dd{p}}{\rho}. 
\Eeq
The differential form of \eq{Enthalpy 2nd Law} suggests that the internal enthalpy depends naturally on the pressure and entropy. 

 If using $h(p,s)$, rather than $e(\rho,s)$, and assuming $\{\rho,s\}$ evolve according to \eqs{rho-eq}{s-eq}, then determining the pressure requires implicitly solving 
\Beq
\label{rho(p,s)-eq} \inv{\rho} =  \thermod{h(p,s)}{p}{s}.
\Eeq
While the distinction between internal energy and enthalpy appears like a mere reorganization of the thermodynamic variables, using the enthalpy and pressure significantly aids in interpreting low Mach number approximations to dynamical models.  

Finally, in addition to \eqs{p(rho,s)-eq}{rho(p,s)-eq}, \eqss{rho-eq}{2nd Law} also imply an explicit local pressure evolution equation.  In differential form
\Beq
\label{dp-eq} 
\dd{p} \ = \ \thermod{p}{\rho}{s} \dd{\rho} + \thermod{p}{s}{\rho} \dd{s}.
\Eeq
Therefore, interpreting the differentials as convective time derivatives produces, 
\Beq
\label{p-eq} && \left(\pd{}{t} + \u \dot\grad{\, }\right)p + \Gamma_{1}(p,\rho) p \,\div{\u } \ = \ 0,
\Eeq
where 
\Beq
\label{Gamma-1-eq} \Gamma_{1}(p,\rho) \equiv \frac{\rho}{p}\thermod{p}{\rho}{s}
\Eeq
specifies the first adiabatic exponent.   
The adiabatic sound speed also follows from \eq{Gamma-1-eq},
\Beq
c^{2}(p,\rho) \equiv \thermod{p}{\rho}{s} = \Gamma_{1}(p,\rho) \frac{p}{\rho}.
\Eeq

In Part~I, we restricted our analysis to a monatomic ideal gas with $\Gamma_{1} = \gamma = 5/3$.  Here, we derive a general sound-proof hydrodynamic model with a general equation of state.  Using general equations of state provides a two-fold advantage over the ideal gas model.  First, many astrophysical application simply require more complex physics than ideal gases can model.  In Appendix~\ref{sec: radiative hydro} we give an example of an astrophysically relevant equation of state incorporating an ideal gas and blackbody radiation, with the general form for $\Gamma_{1}$ given in \eq{eq:gamma1 rad hydro}.  
This is of particular interest when studying convection in massive stars, where near-Eddington luminosities lead to strongly interacting mixtures of radiation and matter, but the fluid dynamics remain very subsonic \citep{Cantiello&Braithwaite_2011}.
One can find these and other examples, including partially ionized gasses critical for stellar interiors, in textbooks on stellar structure \citep[e.g.,][Chapter 9]{Cox&Giuli_stellar_structure}.  Second, keeping a general equation of state helps clarify and unify the thermodynamic interpretation of a number of concepts that arise when considering the energetics of waves and instabilities in stratified atmospheres.   We also show in Appendix~\ref{sec: radiative hydro} how to implement the GPI \eqss{eq:new-momentum}{eq:new-pressure} using this more complex equation of state. 
 
One of the main themes of this paper and Part~I focuses on the consequences of gravitational stratification for low Mach number dynamics.  Therefore, we introduce a hydrostatically balanced, stratified reference atmosphere with background density $\rho_{0}$, pressure $p_{0}$, temperature $T_{0}$, and entropy $s_{0}$ that vary only in the direction of gravity
\Beq\label{eq: hydrostatic}
\grad{p_{0}}  = -\rho_{0}\,  \grad{\phi}.
\Eeq
The potential function $\phi(\vec{x})$ determines the local gravitational acceleration $\g = - \grad{\phi}$.  

Part~I discusses anelastic models, which eliminate acoustic modes from the dynamics by employing the density constraint 
\Beq
\label{anelastic constraint} |\rho - \rho_{0}| \ll \rho_{0} \implies \div{( \rho_{0} \u ) }  = 0.  
\Eeq 
\Eq{anelastic constraint} rests on the premise that dynamical density fluctuations remain small compared to the background.  This assumption indeed filters sound modes, but is only formally justified for nearly adiabatically stratified atmospheres. 

Beginning with \citet{Durran_1989}, a number of authors have recognized that the pressure field more naturally distinguishes between acoustic and low Mach number dynamics.  The pressure field provides the restoring force for acoustic oscillations, and this implies that small pressure fluctuations,
\Beq
| p - p_{0} | \ll p_{0},
\Eeq	
characterize low Mach number dynamics on small scales, rather than small density fluctuations as in \eq{anelastic constraint}.    

Therefore, replacing the pressure field with the static background reference field,
the pseudo-incompressible models in this paper replace the pressure evolution \eq{p-eq} with a velocity divergence equation
\Beq
\label{p0-eq} && \div{\u } +  \frac{\u \dot\grad{p_{0}}}{\Gamma_{1}(p_{0},\rho) p_{0} }  \ = \ 0.
\label{eq:PI continuity}
\Eeq
Moreover, for the case of constant $\Gamma_{1} = \gamma$, we may rewrite \eq{p0-eq} into a form similar to \eq{anelastic constraint}
\Beq
\label{psi-div} \div{\left( \beta_{0} \u \right)} = 0,
\Eeq
where
\Beq
\label{beta-def} \beta_{0} \equiv  p_{0}^{1/\gamma}.
\Eeq 

We consider \eq{p0-eq} the
defining characteristic of pseudo-incompressible models --- as opposed to anelastic models.      
In adiabatically stratified atmospheres, $s_{0} = \mbox{\textit{constant}}$ implies $p_{0} \propto \rho_{0}^{\gamma} $, and
\Beq
  \div{\left(\beta_{0} \u \right)} \propto \div{\left(\rho_{0} \u \right)} = 0.
  \label{eq:PI-Anelastic}
\Eeq
In these atmospheres, the pseudo-incompressible constraint \eq{p0-eq}
reduces to the more common anelastic constraint, which we studied in Part~I.

Here we consider two different pseudo-incompressible models.  The different notation and different thermodynamics used in the various treatments can lead to ambiguity over the equivalence or differences between models.  To avoid this, as in Part~I, we write each model using as consistent a notation as
possible.  Practical \emph{numerical} or computational differences can
arise when \emph{solving} different transformations of the same fundamental model, but these issues mostly lie beyond our current scope.  Therefore, we consider two models identical if one can bring them into the same form by legitimate mathematical transformation (without approximation), and/or by a possible change of notation.  

\subsection{Fully compressible Euler equations \label{sec: Fully compressible Euler equations} }

The thermodynamic \eqss{rho-eq}{p-eq} apply with respect to an arbitrary flow field, $\u (t,\vec{x})$. For fully compressible dynamics neglecting dissipation and sources of heat, the fully compressible Euler (FC) equations govern the flow evolution,
\Beq
\label{u-eq} &&  \rho \left(\Dt{}\right)\u + \grad{p} + \rho \grad{\phi} \ = \ 0
\label{eq:compressible momentum} \n
&&\pd{\rho}{t} + \div{\left(\rho \u\right)} \ = \ 0\n
&&\label{eq:compressible entropy}\Dt{s} \ = \ 0.
\Eeq
An equation of state $p=p(\rho,s)$ closes \eqss{u-eq}{eq:compressible entropy}.
In this paper, with conservation of energy a primary theme, we limit our analysis to strictly isentropic flow, $\!\dd{s} = 0$ (\eq{eq:compressible entropy}).  This allows us to ensure that no mechanical violations of energy balance occur.  Our more general interest lies in proper energy budgets for non-isentropic flows, but including dissipative terms would complicate the analysis presented here, and we intend to consider such effects in future work.  

\Eqss{eq:compressible momentum}{eq:compressible entropy} imply the conservation of total energy 
\Beq
\label{Compressible Energy} \pd{E}{t} + \div{\left[\u (E + p)\right] } \ = \ 0
\Eeq
where 
\Beq
\label{Total Euler Energy} \quad E = \frac{\rho |\u |^{2}}{2} + \rho e(\rho,s) + \rho\phi.
\Eeq  
The FC equations conserve energy with an arbitrary equation of state, assuming $\partial_{t} \phi = 0$.  

The form of the potential energy in \eqs{Compressible Energy}{Total Euler Energy} leads to some confusion when comparing linear and nonlinear dynamics.  For nonlinear states, the potential energy appears to determine very little about how the system may or may not extract energy from a background stratification and convert it into fluid motion.  Alternatively, by perturbing  about a background hydrostatic stratification, the equations of motion possess a well-known energy-conservation principle that adopts a quadratic form in the perturbation variables.  
 
We reconcile these facts by noting that \eqs{rho-eq}{s-eq} together imply
\Beq
\label{Casimir} \pd{\left(\rho f(s)\right)}{t} + \div{\left[\rho \u f(s) \right] } = 0,
\Eeq
\textit{for any} arbitrary function exclusively depending on entropy, $f(s)$. 

\Eq{Casimir} implies the existence of a family of conserved free energies,
\Beq
F \equiv E - \rho f(s)  = \frac{\rho |\u |^{2}}{2} + \rho \left[e(\rho,s) - f(s) + \phi\right]\!,\quad
\Eeq
where we subtract an arbitrary ``ground-state" energy, represented by $f(s)$.   

Given a hydrostatically balanced background stratification $\{\rho_{0}(\vec{x}),p_{0}(\vec{x}),s_{0}(\vec{x})\}$, we may choose $f(s)$ such that the linear contribution from perturbations to the free energy cancel identically.  The character of the quadratic terms then determines the linear stability, and possibly even the nonlinear stability. 

The idea of available potential energy goes back to \cite{Lorenz_1955} in the case of a Boussinesq fluid, and \cite{Andrews_1981} computed the quantity for a compressible system by analyzing the dynamical equations directly.  Using the potential energy, we show in Appendix~\ref{sec: Free Energy} that the functional relations
\Beq
\label{ground state} f(s_{0}(\vec{x})) = h_{0}(\vec{x}) + \phi(\vec{x}), \ \  f^{\prime}(s_{0}(\vec{x})) = T_{0}(\vec{x}),\quad
\Eeq
define the appropriate choice of ground state energy density required to cancel contributions from linear perturbations.  Mathematically, $h_{0} + \phi$, represents the ``pullback'' of $f$ with respect to $s_{0}$.  We may use \eq{ground state} to find a global function, $f(s)$, if $s_{0}$ is a one-to-one function of the gravitational potential, $\phi$. 

\Eq{ground state} renders $F$ quadratic to lowest-order in the perturbations around background state.  That is,
\Beq\label{free energy}
F = \frac{\rho |\u |^{2}}{2} + A - p_{0}(\vec{x}),
\Eeq
where $A$ represents the available potential energy. To leading order,
\Beq
\label{APE} && A \equiv \rho \left( e(\rho,s) + \phi(\vec{x}) - f(s) \right) + p_{0}(\vec{x})\quad \n &&
=
\label{linear-APE} \frac{(p-p_{0})^{2}}{2 \rho_{0} c_{0}^{2}} +  \frac{\rho_{0} N_{0}^{2}(s-s_{0})^{2}}{2 | \grad{s_{0}}|^{2}}+ \mathrm{h.o.t.},
\Eeq
where, 
\Beq
\label{BVF} N_{0}^{2} \equiv \g \dot \left(\frac{\grad{\rho_{0}}}{\rho_{0}} - \frac{\g }{c_{0}^{2}}  \right)
\Eeq
defines the background Brunt-V\"ais\"al\"a frequency, and $c_{0}$ represents the background sound speed.  For linear dynamics, we may neglect the higher-order terms in \eq{APE}, and obtain the well-known quadratic energy invariant in \eq{linear-APE}.  In this case, we see clearly that $N_{0}^{2} > 0$ implies $F \ge -p_{0}$, and the system cannot create kinetic energy indefinitely.  We also see that, to leading order, pressure perturbations always produce a positive contribution to the free energy.  

Additionally, we show in Appendix~\ref{sec: General Buoyancy Frequency} that one may rewrite \eq{BVF} into the alternative general form
\Beq
\label{entropy N_0} N_{0}^2=-\frac{T_{0} \alpha_{T}}{c_{p}}\g \dot\grad s_{0},
\Eeq
where $c_{p}$ gives the specific heat capacity at constant pressure, and $\alpha_{T}$ gives the thermal expansion coefficient at constant pressure,  
\Beq
c_{p} \equiv T \thermod{s}{T}{p},\quad  \alpha_{T} \equiv  - \inv{\rho} \thermod{\rho}{T}{p}.
\label{eq:c_p and alpha_T}
\Eeq
\Eq{entropy N_0} relates entropy gradients to the potential density gradient in \eq{BVF}, and simplifies to the more typical expression in the case of an ideal gas where, $T \alpha_{T} = 1$, and $c_{p}$ is constant.  

Along with energy, entropy and mass, non-dissipative flow dynamics conserves the Ertel potential vorticity 
\Beq\label{eq:q def}
q \equiv \frac{\grad{s} \dot \curl{\u } }{\rho},
\Eeq
such that,
\Beq
\label{PV-conserve} \Dt{q} = 0.
\Eeq
\Eq{PV-conserve} represents a fundamentally important result in continuum fluid mechanics and constrains many important processes.  Conservation of potential vorticity finds many useful applications across geophysical and astrophysical fluid dynamics \citep{Bretherton_1970,Salmon_1988} and neglecting \eq{PV-conserve} admits similar errors as neglecting energy conservation in \eq{Compressible Energy}.

\subsection{Pseudo-incompressible equations}
\label{sec: Pseudo-incompressible equations}

\citet{Durran_1989} first proposed pseudo-incompressible models as a response to inadequacies of anelastic models. As Part~I discusses in detail, anelastic models faithfully describe nearly adiabatic convection zones whilst filtering fast acoustic motions from the dynamics.  Additionally, different anelastic models perform with differing results in stably stratified zones.

The anelastic equations filter sound waves with a condition equivalent to requiring the density field to instantaneously match the background stratification.  As we discuss in the introduction, however, local pressure equilibrium proves the more physically well-motivated assumption when wishing to sound-proof a dynamical model.  The distinction between density and pressure in  approximations becomes especially clear for atmospheres with significant sub-adiabatic (stable to convection) stratification.  In the Sun, this particular difficulty arises in modeling the transition from the convective outer envelope to the stable radiative interior.  This region of penetration and overshoot, known as the tachocline, likely plays an important role in the solar dynamo.  Understanding the coupled dynamics between convection and gravity waves in this region is vitally important.  Sufficiently deeply in the solar radiative zone, the assumption of near adiabaticity underlying the anelastic approximation eventually breaks down, even though the flow remains in a significantly low Mach number regime.  

The work of \citet{Durran_1989}  sought to overcome the difficulty of transitions to strong stable stratifications when modeling similar situations in earth's troposphere and stratosphere.  The main innovation in that work assumed rapid pressure, rather than density, equilibration.  However, even though this model, dubbed the pseudo-incompressible (PI) equations, significantly generalize the anelastic equations, it still implicitly assumes an ideal gas equation of state.  This assumption leaves open the question of how to generalize the idea of pressure balance to the more complex situations encountered in many astrophysical applications.   In this section, we outline the basic assumptions of the PI equations with an ideal equation of state.  After establishing the basic idea, we use a general variational framework (\sec{Lagrangian Analysis}) to move beyond simple equations of state whilst retaining the appealing features of the PI equations in \sec{Constrained Models}.  

The FC equations from \sec{Fully compressible Euler equations} require adjustment to correctly accommodate the pressure-restricted divergence \eq{psi-div}. The PI equations evolves according to
\Beq
 &&\rho\frac{D \u }{Dt}
  = -\beta_{0}\grad{\left(\frac{p^{\prime}}{\beta_{0}}\right)}  + (\rho - \rho_0) \g ,
 \label{eq:PI momentum} \n 
 &&\pd{\rho}{t} + \div{\left(\rho \u\right)} \ = \ 0\n 
 && \div{\left( \beta_0 \u \right)} \ = \ 0 \label{eq:PI constraint} 
\Eeq
where $D/Dt \equiv \partial_t + \u \dot \grad{}$,  and $p^{\prime} = p - p_{0}$ represents the pressure fluctuations around a given hydrostatic background.   
We recall the definition $\beta_0\equiv p_0^{1/\gamma}$ (\eq{beta-def}).  \Eqss{eq:PI momentum}{eq:PI constraint} assume an ideal gas with constant ratio of specific heats $\gamma$.  
Ostensibly, the pressure-gradient term in the momentum \eq{eq:PI momentum} takes an unusual form when compared to the FC momentum \eq{eq:compressible momentum}.  \citet{Durran_1989} originally derived \eq{eq:PI momentum} by writing the FC momentum equation in terms of the Exner function and potential temperature respectively, 
\Beq
\pi \equiv \left(\frac{p}{p_{\rm ref}}\right)^{\frac{\gamma-1}{\gamma}}, \quad \theta \equiv \frac{T}{\pi}
\Eeq
where $p_{\rm ref}$ is a constant reference pressure. The only approximation comes from assuming small Exner function fluctuations in the pressure evolution equation.  \footnote{Transforming the equations in \citet{Durran_1989} into \eqss{eq:PI momentum}{eq:PI constraint} requires identifying $\rho^{*}$ with the mass density $\rho$.} 
As we show in \sec{Constrained Models}, the treatment of pressure in momentum \eq{eq:PI momentum} provides exactly the correct behavior given the pseudo-incompressible divergence constraint.  
\Eqss{eq:PI momentum}{eq:PI constraint} also conserve entropy, assuming an equation of state of the form $s = s(p_{0},\rho)$.  The most important physical assumption of the PI equations is that they ignore pressure fluctuations everywhere except the momentum equation.

The conservation of energy gives one important self-consistent feature of the PI equations,
\Beq
\label{eq:PI energy equation}
 \pd{E_{\mathrm{PI}}}{t}+\div\left[ \u \left(E_{\mathrm{PI}}+p'+\frac{\gamma}{\gamma-1}p_0\right)\right]=0,
 \Eeq
where,
\Beq\label{eq:PI energy}
E_{\mathrm{PI}} \equiv \frac{\rho|\u |^2}{2}+\rho\phi.
\Eeq 
We note that $\gamma p_0/(\gamma-1)=\rho_{0} h_0 $ coincides with the background enthalpy density for an ideal gas.  \Eq{eq:PI energy equation} states that filtering acoustic motions from the dynamics should not produce energy balance violations in low Mach number flow.  We expand on the physical reasons for this important fact in \sec{Constrained Models}.

Similar to the FC equations, the PI equations also conserve a free energy that assumes a quadratic form for small perturbations, 
\Beq
\label{linear-free-energy}  F &=&  \rho_{0}\frac{|\u|^{2}}{2} + \frac{\rho_{0} N_0^2 (s-s_{0})^{2}}{2|\grad s_{0}|^{2}}+ \mathrm{h.o.t.}  \n  \nonumber & = & \rho_{0}\frac{|\u|^{2}}{2} + \frac{g^2}{N_0^2}\frac{(\rho-\rho_{0})^{2}}{2\rho_0}+ \mathrm{h.o.t.},
\Eeq
where we only retain the leading-order quadratic form for linear wave dynamics.  Importantly, \eq{linear-free-energy} only omits free potential energy associated with acoustic oscillations.  Dropping pressure dependencies allows us to use only the density perturbations, for sources of potential energy. 
\Eq{ground state} determines the adjustment, $f(s)$, to \eq{eq:PI energy} that produces \eq{linear-free-energy}, just as for the FC equations.  The conservation of \eq{linear-free-energy} for small-amplitude gravity waves makes an important point of comparison to other models in \sec{numerical}.

In addition to properly conserving energy, the PI equations also consistently account for potential vorticity in the reduced dynamics.  In this situation, the entropy gradient reflects the fact that $p \approx p_0$,
\Beq\label{eq:PI grads}
\grad s = c_p \left(\frac{\grad p_0}{\gamma p_0}-\frac{\grad \rho}{\rho}\right)
\Eeq
and
\Beq
\left(\Dt{}\right)\left(\frac{\grad{s} \dot \curl{\u } }{\rho}\right) = 0.
\Eeq

\subsection{MAESTRO low Mach number equations}
\label{sec: MAESTRO low Mach number equations}

In the last few years, a new subsonic model came into the astrophysical community from work modeling Type-Ia supernovae progenitors.  These low-Mach number (LM) equations , resemble the PI equations but with the important feature of a generalized equation of state needed to model the complex thermodynamics of convecting white-dwarf interiors \ann.  
The MAESTRO code implements the LM equations in an adaptive resolution framework suited to the needs of multiscale convection problems
\citep{Zingale_et_al_2009, 
Nonaka_et_al_2010, 
Nonaka_et_al_2012, 
Malone_et_al_2011}.

The LM equations use the same momentum and density equations as the standard-form
Euler \eqs{rho-eq}{u-eq}.  However, the velocity-divergence \eq{p0-eq} replaces the pressure \eq{p-eq}.  We write the LM equations set as
\Beq
&&  \rho \frac{D\u }{Dt} \ = \ -\grad{p'}  +  (\rho - \rho_0) \g 
 \label{eq:LM momentum} \n
&&\pd{\rho}{t} + \div{\left(\rho \u \right)} \ = \ 0 \label{eq:LM continuity} \n 
&& \div{\u } +  \frac{\u \dot\grad{p_{0}}}{\Gamma_{1}(p_{0},\rho) p_{0} }  \ = \ 0, \label{eq:LM divergence}
\Eeq
where we again use $p'$ to denote the pressure perturbation.  We here exclude many dissipative and multi-species effects that are important in the modeling of dynamics in actual white-dwarf interiors; including these effects would not change the fundamental conclusions we draw here about the LM equations.

The LM momentum \eq{eq:LM momentum} is motivated by scaling the FC momentum equation (written in terms of $p$ and $\rho$) with respect to the characteristic velocity $U_{\rm ref}$, the timescale $t_{\rm ref}=U_{\rm ref}/g$, and the length scale $L_{\rm ref}=U_{\rm ref}/t_{\rm ref}=U_{\rm ref}^2/g$ \citep{Almgren_et_al_2006a_ApJ}.  The density and pressure scale with respect to $\rho_{\rm ref}$ and $p_{\rm ref}$, and the characteristic pressure scale height $H_{\rm ref}=p_{\rm ref}/(\rho_{\rm ref}g)$ is defined such that $L_{\rm ref}/H_{\rm ref}=\O{\Ma^2}$.  These choices imply that the non-dimensional momentum equation becomes
\Beq
\label{eq:LM motivation}
\rho_{*} \frac{D \u_{*} }{Dt_{*}}  = -\inv{\Ma^2}\grad_{*}(p_{*}-p_{0,*})  +  (\rho_{*} - \rho_{*,0}) \g_{*},\quad
\Eeq
where the $*$ subscripts represent non-dimensional scaled quantities.  
The non-dimensional pressure fluctuations must remain $\mathcal{O}(\Ma^{2})$ in order to balance time evolution and buoyancy, and omitting the evolution of these fluctuations from \eq{p-eq} filters acoustic modes and allows for volume changes by moving through the background stratification \citep{Almgren_et_al_2006a_ApJ} .  However, \eq{eq:LM motivation} gives a different momentum equation than \eq{eq:PI momentum} because transforming the original thermodynamic variables in the FC equations, and then approximating the result, gives a different result than making the approximations first and then transforming to new variables.  This particular derivation neglects possible changes in buoyancy due to momentum exchange --- but not energy exchange --- between the low Mach number flow and rapid sound waves.

In trying to derive an energy conservation principle for the LM equations, the following form gives the closest possible analogue to \eqs{Compressible Energy}{eq:PI energy equation},
\Beq\label{eq:LM energy}
\pd{E_{\mathrm{LM}}}{t} +\div\left( \u E_{\mathrm{LM}}\right) = -\u \dot\grad p',
 \Eeq
where 
\Beq
E_{\mathrm{LM}}=\frac{\rho|\u |^2}{2}+\rho\phi+\rho h(p_0,s)=E_{\mathrm{PI}}+\rho h(p_0,s),\quad
\Eeq
and $h(p_0,s)$ evolves according to \eq{Enthalpy 2nd Law}.  The term on the right-hand side of \eq{eq:LM energy} does not reduce to an exact divergence, and represents an uncontrolled energy source or sink.  For an ideal gas with constant ratio of specific heats, $\gamma$, \eq{eq:LM energy} becomes
\Beq
 \pd{E_{\mathrm{PI}}}{t}+\div\left[ \u \left(E_{\mathrm{PI}}+\frac{\gamma}{\gamma-1}p_0\right)\right]=-\u \dot\grad p',\quad
\Eeq
which only differs from \eq{eq:PI energy equation} by $p'\div{u}$ on the right-hand side.  
For general flows, the LM equations do not conserve any energy $E$.

Along with energy, the LM equations also do not conserve potential vorticity.  To show this, we use \eq{ds-dp-drho} 
\Beq\label{eq:grads}
T \alpha_T \frac{ds}{c_p} =  \frac{dp}{\Gamma_1 p}-\frac{d\rho}{\rho}, \nonumber
\Eeq
where $\alpha_T$ is the coefficient of thermal expansion given in \eq{eq:c_p and alpha_T}.  Because of the pressure constraint, $p \approx p_{0}$, the entropy for the LM equations only depends on density $\rho$ and $p_{0}$, thus $s = s(p_0(\vec{x}),\rho)$.  The entropy $s$ still satisfies \eq{s-eq}.
Therefore, $q=\grad s \dot(\curl\u )/\rho$ still defines the potential vorticity, using
\Beq\label{eq:LM grads}
\grad s = \frac{c_p}{T\alpha_T}
 \left(\frac{\grad p_0}{\Gamma_1 p_0}-\frac{\grad\rho}{\rho}\right),
\Eeq
where $c_p, T, \alpha_T$, and $\Gamma_1$ all implicitly depend on $p_0$ and $\rho$.
Therefore,
\Beq
\label{eq:LM PV}
 \Dt{q} = \frac{c_p}{\rho^3T\alpha_T\Gamma_1 p_0} \grad p_0\dot\left(\grad\rho\cross\grad p'\right).
\Eeq
If we assume an ideal gas equation of state, then \eq{eq:LM grads} reduces to \eq{eq:PI grads} and \eq{eq:LM PV} reduces to
\Beq
 \Dt{q} = \frac{c_v}{\rho^3p_0} \grad p_0\dot\left(\grad\rho\cross\grad p'\right).
\label{eq:LM PV ideal gas}
\Eeq
No way exists to eliminate the right-hand side of equations~(\ref{eq:LM PV}) or (\ref{eq:LM PV ideal gas}).  The LM equations do not conserve potential vorticity for general motions.

Problems remain even in the linear regime.   The right-hand side of \eq{eq:LM PV} does vanish if we consider only linear perturbations, so linear perturbations do conserve potential vorticity, which vanishes for gravity waves.  They do not however conserve energy. In the linear regime, the conservation of $s$ and the definition of the buoyancy frequency in \eq{entropy N_0} implies
\Beq
\label{LM-linear-E} \pd{E}{t}+\div\left(\u p'\right) = -p' \frac{\u \dot\grad p_0}{p_0\Gamma_1(p_0,\rho_0)},
\Eeq
where $E$ follows from \eq{linear-free-energy} just as in the PI equations.
\Beq
&& E = K + U \label{linear-energy}, \\
&& K = \frac{\rho_0|\u |^2}{2} \label{linear-kinetic energy}, \\
&& U =  \frac{g^{2} (\rho - \rho_{0})^{2}}{2N_0^2 \rho_0}, \label{linear-potential energy}
\Eeq
where we use the density, rather than entropy, form of \eq{linear-free-energy}.
Even for linear motions the right-hand side of \eq{LM-linear-E} remains non-zero.

Although the LM equations do not conserve energy, the linearized equations do possess a quadratic invariant, that we call the pseudo-energy.  As in Part~I, the presence of a pseudo-energy can produce misleading effects since it allows bounded solutions in some dynamical regimes.  However, bounded solutions do not necessarily imply accurate solutions, and as we found in Part~I, these pseudo-energy solutions may contain significant errors in many situations.  We now derive the conserved pseudo-energy.  

Assuming $p_0$ and $\rho_0$ only depend on the direction of $\vec{g}$, which we call $z$, we follow \citet{Almgren_et_al_2006a_ApJ} and define a more general 
\Beq
\label{general-beta0}\log\beta_0 \equiv \int\! \frac{\dd{p_0}}{p_0\Gamma_1(p_0,\rho_0)}=  - \int \!\frac{g \dd z}{c_{0}^{2}}.
\Eeq 
For the simple case of an ideal gas with constant $\gamma$, \eq{general-beta0} recovers $\beta_0=p_0^{1/\gamma}$.
Using the general form of $\beta_0$ in \eq{LM-linear-E} produces
\Beq
\pd{\PE}{t}+\div(\u p'\beta_0)=0,
\label{LM-linear-PE}
\Eeq
where $\PE$ represents the quadratic invariant  pseudo-energy
\Beq\label{pseudo-energy}
&& \PE \equiv \beta_0E = \PK + \PU , \n
&& \PK \equiv \beta_0K=\frac{\rho_0\beta_0|\u |^2}{2} \label{pseudo-kinetic energy} , \n
&& \PU \equiv \beta_0U=\frac{g^{2}\beta_{0} (\rho - \rho_{0})^{2}}{2N_0^2 \rho_0}, \label{pseudo-potential energy}
\Eeq 
\Eq{LM-linear-PE} implies that linear motions in the LM equations conserve pseudo-energy $\PE$.

The pseudo-energy and energy only coincide for constant $\beta_{0}$, which implies an atmosphere with no background stratification.  For an adiabatic atmosphere $\beta_{0} \propto \rho_{0}$, and the LM do not reduce to anelastic equations.   
In \sec{numerical}, we numerically calculate the evolution of the energies (\eqss{linear-energy}{linear-potential energy}) and pseudo-energies (\eqss{pseudo-energy}{pseudo-potential energy}) for a simple test problem with both the PI and LM equations.

\section{Lagrangian Analysis}
\label{sec: Lagrangian Analysis}

The analysis in \sec{Background and model equations} makes clear that the form of the momentum equation determines if a given sound-proof model will correctly conserve energy.  Generally, as we found in Part~I for anelastic models, sound-proof equations using an unmodified Euler momentum equation do not conserve energy.  In this section we develop the tools which allow one to correctly and consistently determine the momentum equation for a soundproof model with general equations of state.

\subsection{Eulerian Action Principle}
\label{sec: Eulerian Action Principle}

Students in fluid mechanics commonly first learn to derive \eq{u-eq} by appealing to Newton's laws of motion for a continuum media \citep{Landau&Lifshitz, Kundu&Cohen}.  This type of analysis typically considers fluid elements either moving with the flow (Lagrangian formulation), or in fixed volumes (Eulerian formulation) and analyzes the forces acting on these fluid elements.  Given the analysis of forces, typically both of these approaches fall under the scope of Newtonian mechanics, regardless of where one places the spatial coordinates.  Alternatively, Lagrangian mechanics relies on Hamilton's principle of stationary action to reformulate a dynamical equations of motion without recourse to specific forces.  Here we derive the fully compressible Euler momentum \eq{u-eq} using Lagrangian mechanics.

In contrast to Newtonian dynamics, Lagrangian mechanics defines a Lagrangian function $\mathcal{L}$,  which is related to the energy at each point of the medium.  The action, $\mathcal{S}$, is the space-time integral of the Lagrangian density.  Hamilton's principle states that the equation of motion is equivalent to the stationarity of the action with respect to the variables of the system.

Hamilton's principle assumes a simple form and interpretation when considering particle mechanics.  The situation in continuum fluid mechanics is somewhat more complicated.  In our case, the problem corresponds to the stationarity of the action
\Beq
\mathcal{S} \equiv \int \mathcal{L}(\u ,\rho,s) \dV \dt,
\Eeq
such that the density and entropy satisfy \eqs{rho-eq}{s-eq}.  We leave the particular form of the Lagrangian unspecified until the next section. Given the similarities with particle mechanics, the Lagrangian coordinate formulation of a fluid lends itself well to solution by Hamilton's principle \citep{Salmon_1988}.  In this case, after extremizing the action, one must take the trouble of recasting the dynamic equations into an fixed coordinate (Eulerian) form to compare with the Newtonian-derived \eq{u-eq}.  Alternatively, one may consider an Eulerian version of Hamilton's action principle on a fixed spatial coordinate system.  In this case, the added difficulty results from properly enforcing the geometric conservation constraints of mass and entropy \eqs{rho-eq}{s-eq}.  

In Eulerian coordinates, the volume measure remains fixed.  Therefore, Hamilton's principle becomes    
\Beq
\label{Action-eq} \delta \mathcal{S}  =  \int \! \!\left(\pd{\mathcal{L}}{\u}\dot \delta \u + \pd{\mathcal{L}}{\rho}\delta\rho + \pd{\mathcal{L}}{s} \delta s\right)\!\!\dV \dt  = 0.\quad
\Eeq
Two constraints relate the virtual displacements in \eq{Action-eq},
\Beq
\label{delta rho-eq} &&\pd{\delta \rho}{t} + \div \left( \rho \delta \u +  \u \delta \rho \right) \ = \ 0 \n
\label{delta s-eq} && \Dt{\delta s} + \delta \u \dot \grad{ s } \ = \ 0.  
\Eeq
\Eqs{delta rho-eq}{delta s-eq} follow by linearizing \eqs{rho-eq}{s-eq} around a given phase-space trajectory $\{\u (t,\vec{x}),\rho(t,\vec{x}),s(t,\vec{x})\}$.  Two main methods exist for enforcing \eqs{delta rho-eq}{delta s-eq} in Eulerian variables.  The first method uses Lagrange multipliers to constrain the action directly. This method of Lin constraints, requires the added complication of Euler (a.k.a.~Clebsch) potentials for the velocity \citep{Cendra&Marsden_1987}.  The second method for enforcing \eqs{rho-eq}{s-eq} amounts to directly parameterizing \eqs{delta rho-eq}{delta s-eq}.  A well-known result from linear analysis, the following perturbations
\Beq
\label{delta-u} &&\delta \u \ = \ \Dt{\vxi} - \vxi \dot \grad{\u }\n
\label{delta-rho} && \delta \rho \ = \ - \div{\left(\rho \vxi\right)}\n
\label{delta-s} && \delta s \ = \ -\vxi\dot \grad{s}
\Eeq
directly solve \eqs{delta rho-eq}{delta s-eq} for any displacement
$\vxi$, and any flow field $\u $ assuming $\rho$ and $s$
satisfy \eqs{rho-eq}{s-eq} \citep{Eckart_1960,Newcomb_1962,Kulsrud}.   The three-dimensional infinitesimal displacement, $\vxi$, generates the five-dimensional constrained phase space.  One may easily confirm \eqss{delta-u}{delta-s} if $\vxi = \u \delta t$, $\delta \u = (\partial_{t} \u) \delta t$, $\delta \rho = (\partial_{t} \rho) \delta t$, and $\delta s = (\partial_{t} s) \delta t$. For this special case, \eq{delta-u} reduces to a trivial statement, and \eqs{delta-rho}{delta-s} become \eqs{rho-eq}{s-eq} respectively.  

Using \eqss{delta-u}{delta-s} in \eq{Action-eq} and integrating by-parts to isolate $\vxi$ gives the Euler--Lagrange equations for an ideal fluid in a fixed spatial Cartesian coordinate system,
\Beq
\label{Euler--Lagrange} && \nn \pd{}{t}\!\! \left( \pd{\mathcal{L}}{u_{i}} \right) + \pd{}{x_{j}}\! \! \left(u_{j} \pd{\mathcal{L}}{u_{i}} \right) + \left( \pd{\mathcal{L}}{u_{j}}\right)\!\pd{u_{j}}{x_{i}}\n
&& - \rho  \pd{}{x_{i}}\!\! \left( \pd{\mathcal{L}}{\rho} \right) + \left( \pd{\mathcal{L}}{s}  \right)\!\pd{s}{x_{i}} \ = \ 0,
\Eeq
where we use Einstein notation and sum over repeated subscripts.  

\Eq{Euler--Lagrange} represents the Euler--Lagrange equations of motion for a general non-dissipative fluid.  At this stage, we specify nothing about the specific form of the Lagrangian density function.  This master equation allows us to derive the FC momentum \eq{u-eq} from an alternative perspective, but more importantly allows us to make approximations in a general way and distill dynamical models that remain consistent with regard to energy, and potential vorticity budgeting, as we now demonstrate.

\subsection{Energy conservation}
\label{sec: Energy conservation}

We define the momentum density
\Beq
\vec{j} \equiv \pd{\mathcal{L}}{\u }.
\Eeq
With this, conservation of energy follows by contracting \eq{Euler--Lagrange} with the velocity field, $\u $,  
\Beq
\nn &&\u \dot \pd{\vec{j}}{t} + \div{\left[\u \, (\u \dot \vec{j}) \right]} - \rho \u \dot \grad{\pd{\mathcal{L}}{\rho}} + \u \dot \grad{s} \pd{\mathcal{L}}{s} =  \quad \n
&& \pd{(\u \dot\vec{j} - \mathcal{L})}{t} + \div \left[ \u \left( \u \dot \vec{j} - \rho \pd{\mathcal{L}}{\rho}\right) \right] = 0,
\label{eq:Euler-Lagrange energy}
\Eeq
where the last line assumes that the Lagrangian $\mathcal{L}$ does not depend on time explicitly.  We recognize the Legendre transforms of the Lagrangian density with respect to the velocity and density give the ``Hamiltonian" $\mathcal{H}$, and the ``generalized pressure" $\mathcal{P}$, respectively, 
\Beq
\mathcal{H} \ = \ \u \dot \vec{j} - \mathcal{L},\quad
 \mathcal{P} \ = \ \mathcal{L} - \rho \pd{\mathcal{L}}{\rho}.
\label{eq:general H and P}
\Eeq
From \eqs{eq:Euler-Lagrange energy}{eq:general H and P}, the general statement of local energy conservation follows in conservative form,
\Beq
\label{eq: energy conservation} \pd{\mathcal{H}}{t} + \div{\left[\u \left(\mathcal{H} + \mathcal{P}   \right)  \right]} \ = \ 0.
\Eeq
The primary utility of \eqs{Euler--Lagrange}{eq: energy conservation} lies in the unspecified form of the Lagrangian density $\mathcal{L}$.  If we choose the traditional form for $\mathcal{L}$  (\eq{L-density-0}), as we will in \sec{Lagrangian Derivation of the Euler Equations}, then \eq{eq: energy conservation} leads to \eq{Compressible Energy} and we have energy conservation.  Furthermore, just as in \sec{Fully compressible Euler equations} and \eq{Casimir}, we may define the conserved free energy
\Beq
\mathcal{F} \equiv \mathcal{H} - \rho f(s)
\Eeq
for any $f(s)$.  Assuming the following relationship 
\Beq
f(s_{0}(\vec{x})) = \pd{\mathcal{H}}{\rho}\bigg|_{\u=\vec{0},\rho=\rho_{0}(\vec{x}),s=s_{0}(\vec{x})}
\Eeq
eliminates all linear contributions to the available potential energy, thus generalizing \eq{ground state}.  We also encourage the reader to compare the Lagrangian formulation for fluid mechanics with the Hamiltonian formulation \citet{Morrison_1998}.

\subsection{Potential vorticity conservation}
\label{sec: Potential vorticity conservation}

We now turn to the conservation of potential vorticity.  We begin by noting that
for any $\varphi$, if
\Beq
\label{s, rho invariant} \quad  \vxi = \frac{\grad{\varphi} \cross \grad{s}}{\rho},
\Eeq
then $\delta \rho = \delta s = 0$, but $\delta \u \ne 0$.  

The interpretation of potential vorticity becomes clear when one realizes that \eqs{rho-eq}{s-eq} imply that density and entropy act as dynamic coordinate variables.  \Eqss{delta-u}{delta-s} show that $\partial_{t}\vxi$ generates three-dimensional virtual displacement velocity, and $\vxi$ generates density and entropy changes.  However, this only accounts for five out of six possible phase-space dimensions. \Eq{s, rho invariant} therefore implies a displacement direction that generates neither $\delta \rho$, nor $\delta s$, but still produces a momentum.  The Lagrangian density only depends on two thermodynamic variables, and hence displacements associated with \eq{s, rho invariant} represents a so-called ``cyclic coordinate'' familiar from Lagrangian particle mechanics.  Each cyclic coordinate implies the conservation of an associated conjugate momenta.  We show here that because of this invariance, the Lagrangian framework manifestly conserves energy \textsl{and} potential vorticity.  

First we define the following two quantities
\Beq
\vec{U} &\equiv& \inv{\rho}\pd{\mathcal{L}}{\u },\quad \vec{\Omega} \equiv \curl{\vec{U}}
\Eeq
For simple Newtonian dynamics, $\vec{U} = \u $ and $\vec{\Omega}$ represents the traditional vorticity; in a rotating frame $\vec{\Omega}$ also includes contributions from the background ``planetary" vorticity.  
Then defining the following quantity,
\Beq
\vec{A} &\equiv& \pd{\vec{U}}{t} + \vec{\Omega} \cross \u + \grad{\left(\vec{U}\dot \u \right)} ,
\Eeq
variation of the action over $\varphi$ from \eq{s, rho invariant} implies
\Beq
\grad{s}\dot \left(\curl{\vec{A}} \right) = 0.
\Eeq
Defining the generalized potential vorticity 
\Beq
\mbox{GPV} \equiv \frac{\grad{s}\dot \vec{\Omega}}{\rho}
\Eeq
and then combining \eqs{rho-eq}{s-eq} and some vector calculus identities gives conservation of $\mbox{GPV}$ along trajectories of the flow
\Beq
\label{GPV conservation} \left(\Dt{}\right)\left[\frac{\grad{s}}{\rho}\dot \curl{\left(\inv{\rho}\pd{\mathcal{L}}{\u }\right)}\right] \ = \ 0.
\Eeq
\Eq{GPV conservation} (for any $\mathcal{L}$) represents a significant additional advantage of the Lagrangian approach, especially when we begin to apply approximations to the fully compressible dynamics. 

In the case where the flow contains other frozen-in dynamical quantities --- such as magnetism in Appendix~\ref{sec: Constrained Magnetohydrodynamics}, or chemical concentration --- then variation in the direction of \eq{s, rho invariant} produces explicit changes in the Lagrangian density and \eq{GPV conservation} no longer holds. 

\subsection{Lagrangian Derivation of the Euler Equations}
\label{sec: Lagrangian Derivation of the Euler Equations}

With \eq{Euler--Lagrange}, we derive here the FC momentum equation with the Lagrangian density,
\Beq
\label{L-density-0} \mathcal{L} \ = \ \rho\left( \frac{|\u |^{2}}{2} -  e(\rho,s) -  \phi \right),
\Eeq
where \eq{2nd Law} gives $e = e(\rho,s)$. Computing specifically each relevant term 
\Beq
\pd{\mathcal{L}}{\u } &=& \rho \u , \\
\pd{\mathcal{L}}{\rho} &=& \frac{|\u |^{2}}{2}  - e - \frac{p}{\rho} - \phi, \\
\pd{\mathcal{L}}{s} &=& \rho T. 
\Eeq
Substituting into \eq{Euler--Lagrange}, 
\Beq
&&\nn \pd{(\rho \u )}{t} + \div{\left( \rho \u \u \right)} + \rho \frac{\grad{|\u |^{2}}}{2}\n 
\label{conservative-u-eq} &&- \rho\, \grad{}\left( \frac{|\u |^{2}}{2} - e - \frac{p}{\rho} - \phi \right) - \rho T \,\grad{s} \ =\  0.
\Eeq
The last two terms in \eq{conservative-u-eq} simplify because of the Second Law of Thermodynamics (\eq{2nd Law}) 
\Beq
\nn T \grad{s}  = \grad{e} - \frac{p}{\rho^{2}}\grad{\rho},
\Eeq
and the FC momentum \eq{u-eq} follows in compact form
\Beq
\nn \rho \frac{D \u }{Dt} + \grad{p} + \rho \grad{\phi} \ = 0.
\Eeq

For the FC equations, the energy adopts the expected form
\Beq
\nn \mathcal{H} \ \equiv \ \u \dot \pd{\mathcal{L}} {\u } - \mathcal{L}  = \rho\left( \frac{|\u |^{2}}{2} +  e(\rho,s) +  \phi \right).
\Eeq
Furthermore, the Legendre transform of $\mathcal{L}$ with respect to $\rho$ reduces to the thermodynamic pressure 
\Beq
\label{Euler pressure} \mathcal{P} \equiv \mathcal{L} - \rho \pd{\mathcal{L}}{\rho}  = p 
\Eeq

At this point, using Hamilton's principle to derive the momentum equations merely reformulates a well-known equation set.  The utility of \eq{Euler--Lagrange} lies in that we need not use the specific Lagrangian density $\mathcal{L}$ in \eq{L-density-0}.  When considering low Mach number flows we also possess the tools to produce constrained dynamics that manifestly filter acoustic modes.  Moreover, any constrained dynamics derived from \eq{Euler--Lagrange} will automatically achieve proper energy balance; this also leads to correct energy budgeting even if the constraint depends explicitly on time.  

\section{Constrained Models}
\label{sec: Constrained Models}

We now use the Euler--Lagrange \eq{Euler--Lagrange} to derive a momentum equation for subsonic constrained flows.  To obtain constrained dynamics, we first recall the hydrostatic reference state
\Beq
\label{eq:hydrostatic balance}
\grad{p_{0}} + \rho_{0}\grad{\phi} \ = \ 0. \nn
\Eeq
As a fluid parcel moves slowly through a background stratification its internal pressure must equilibrate rapidly to approximately that of its surroundings.  If the parcel does not remain almost in pressure balance then it could isotropically compress or expand without moving vertically.  Rapid compressive collapse and/or expansion generate the fast oscillations characterizing  acoustic motions. We wish to remove these modes from the dynamics.  In this section we do so, and we will see that a succession of more restrictive assumptions leads to the generalized pseudo-incompressible, anelastic, and Boussinesq equations in turn.

\subsection{Lagrangian Derivation of the Generalized Pseudo- Incompressible Equations}
\label{sec: Lagrangian Derivation of the Pseudo- Incompressible Equations}

The first and least restrictive constraint is that pressure
fluctuations equilibrate rapidly and are small.
We therefore impose the following (velocity independent, or holonomic) constraint on the dynamics
\Beq
\label{eq:constraint} \C = p(\rho,s) - p_{0}(\vec{x}) \ = \ 0.
\Eeq
Enforcing pressure equilibrium implicitly requires the fluid to move in some reduced subset of the phase space.  

The example of a roller coaster on a track provides an appropriate analogy to the fluid problem.  
The track (background pressure stratification) is strictly speaking a dynamical system in and of itself, and in particular supports elastic flexural waves (acoustic waves). The roller coaster (subsonic swirling motion) couples to the very stiff (high sound speed) steel rails, which flex and push on the train as it moves along (fast acoustic restoration).  But simply requiring the constraint (pressure equilibrium) says that the steel in the track responds so rapidly, that effectively the train must slide along parallel to the rails. In such a system, the train's momentum is not conserved.  The track must impart momentum to the train, otherwise the roller coaster would merely plunge straight downward (radially divergent pressure-driven collapse).  If the initial energy of the roller car is sufficiently large compared to the potential energy in the rails (high Mach number situation), then the constraint no longer suffices, and the dynamics of the car and track progress in a fashion where it becomes difficult to distinguish between the separate dynamics of the two components.  
Ignoring here the possibility of a catastrophe (high Mach number flows), in a frictionless system energy remains conserved and we may derive the equations of motion via the action principle paired with a Lagrange multiplier and constraint.    

Here we derive equations of motion for low-Mach number subsonic flows using constrained Lagrangian analysis.  For a system under constraint, the Lagrangian density becomes 
\Beq
\label{eq:constrained lagrangian density}
\mathcal{L} \to \mathcal{L} - \lambda\, \C(\rho,s),
\Eeq
where \eq{eq:constraint} gives the constraint $\C$, and $\lambda$ represents the Lagrange multiplier, an additional independent variable of the action.
Thus the full Lagrangian density becomes
\Beq
\label{constrained L-density} \mathcal{L} \ = \ \rho\left( \frac{|\u |^{2}}{2} -  e(\rho,s) -  \phi \right) - \lambda\, \C(\rho,s).
\Eeq
The appropriate new term in the Euler--Lagrange \eq{Euler--Lagrange} assumes that 
stationarity of the action with respect to $\lambda$ enforces the constraint,
\Beq
\label{eq:constrained lagrangian E-L term}
\pd{\mathcal{L}}{\lambda} \ = \ -\C(\rho,s) \ = \ -p + p_{0} \ = \ 0.
\Eeq
Stationarity of the action with respect to $\vxi$ yields the equations of motion as in \eq{Euler--Lagrange}.

The strictly additive constraint in \eq{constrained L-density} implies the portion of the dynamical equations deriving from the original (unconstrained) Lagrangian density remain unaltered.  Therefore, the constrained Euler--Lagrange equations assume the general form 
\Beq
\nn && \rho \frac{D\u }{Dt} + \grad{p} + \rho \grad{\phi} = \n 
\nn && - \rho \grad{\left( \lambda \pd{\mathcal{C}}{\rho}\right)} + \lambda \pd{\mathcal{C}}{s} \grad{s} 
= - \grad{\left(\Gamma_{1} \lambda p \right)} + \lambda \grad{p}, 
\Eeq
where we use
\Beq
\pd{\mathcal{C}}{\rho} = \thermod{p}{\rho}{s},\quad \pd{\mathcal{C}}{s} = \thermod{p}{s}{\rho}.
\Eeq
Defining the rescaled Lagrange
multiplier
\Beq
\label{pi-def} \tilde{p} \equiv \Gamma_{1} p \lambda,
\Eeq
the constrained momentum equation now becomes
\Beq
\label{new-u-eq} \rho \frac{D\u }{Dt} + \grad{p} + \rho \grad{\phi} \ = \ - \grad{\tilde{p}} + \frac{\tilde{p}}{ \Gamma_{1} p }\grad{p}.
\Eeq
Comparing \eq{new-u-eq} with \eq{u-eq}, we see that \eq{new-u-eq} now contains an additional source of momentum that prevents the dynamics from leaving the constraint manifold.

The constraint \eq{eq:constraint} allows us to replace freely $p \to p_{0}$, leading to the following constrained low-Mach number equations for momentum, density, and energy:
\Beq
\label{eq:new-momentum} && \rho \frac{D\u }{Dt} + \left(\rho - \rho_{0}\right) \grad{\phi} = - \grad{\tilde{p}} + \frac{\tilde{p}\, \grad{\ln p_{0}}}{\Gamma_{1}(p_{0},\rho)}, \n
\label{eq:new-density} &&\frac{ D \rho}{Dt}  =  \rho \frac{\u \dot \grad{p_{0} }}{\Gamma_{1}(p_{0},\rho)p_{0}}, \n
\label{eq:new-pressure} && \div{\u }  = - \frac{\u \dot \grad{\ln p_{0} }}{\Gamma_{1}(p_{0},\rho)},
\Eeq
We dub this \eqss{eq:new-momentum}{eq:new-pressure} the Generalized Pseudo-Incompressible (GPI) equations.
The scaled Lagrange multiplier $\tilde{p}$ enforces \eq{eq:new-pressure}.
In \eq{eq:new-momentum} we use \eqs{eq: hydrostatic}{eq:constraint} to replace $\grad{p} \to \grad{p_{0}} = -\rho_{0} \grad{\phi}$, but the generality of \eq{new-u-eq} does not require hydrostatic balance.  In the limit of an ideal gas equation of state with constant $\gamma$, \eqss{eq:new-momentum}{eq:new-pressure} reduce to the PI \eqss{eq:PI momentum}{eq:PI constraint}.  Thus, the GPI equations extend the PI equations to more general equations of state.  

Most previous derivations of  anelastic or pseudo-incompressible models follow from an asymptotic expansion of the full equations under the assumption of particular scalings of the leading-order terms. The asymptotic procedure applies to many systems under strong constraint \citep{Julien&Knobloch_2007}.  \cite{Spiegel&Veronis_1960} pioneered this approach with the Boussinesq approximation for a compressible fluid, \cite{Gough_1969} does this for the anelastic models discussed in Part~I, while \citet{Durran_1989}, \cite{Achatz_et_al_2010}, and \cite{Klein_et_al_2010} do this for the pseudo-incompressible model.  In all of these derivations, the authors implicitly assume $|p-p_{0}| \ll p_{0}$, which may or may not imply a similar relation for density.

Our approach makes the same assumptions as previous authors.  Our method differs in that we enforce the pressure constraint identically; \ie non-perturbatively.  The main disadvantage of our approach is that our constraint tells us nothing quantitative about when our assumptions breakdown.  We must apply physical reasoning (as in the roller coster analogy) and solution monitoring to determine when acoustic and swirling motions loose their distinct identity and thus when our derived equations lose their validity.  
The main advantage of our derivation is that it provides a general, systematic, framework for extending the PI equations to include additional physics.  Hamilton's principle gives a systematic method for generalizing the PI equations to an arbitrary equation of state, while correcting the energy violations that emerge in the LM equations.

We can apply the same Lagrangian framework to other important problems in astrophysical and geophysical fluid mechanics.  For example, we derive an analogous model in Appendix~\ref{sec: Constrained Magnetohydrodynamics} that incorporates magnetism correctly in a general framework.
We note that any shortcomings of the GPI equations derived here also apply to the previous sets of sound-proof equations discussed in \sec{Introduction and motivation}.  In particular, the GPI equations reduce to the PI equations for an ideal gas and to the anelastic equations for an adiabatic background stratification.  
Furthermore, even if a filtered model gives lacklustre \textit{quantitative} results, the equations may still give excellent \textit{qualitative} insight by eliminating complicating details.  Finally, well-motivated subsonic models may guide the construction of numerical methods that take fast dynamics into account in an optimal fashion.

The energy density in the GPI equations retains the same form as for the the FC equations, \ie
\Beq
\mathcal{H} \equiv \u \dot \pd{\mathcal{L}}{\u } - \mathcal{L} = \rho \left( \frac{|\u |^{2}}{2} + e + \phi\right),
\Eeq
where implicitly $p=p_{0}$.

However, we score more physical interpretation when computing the generalized pressure, 
\Beq
\nn && \mathcal{P} \equiv \mathcal{L} - \rho \pd{\mathcal{L}}{\rho} =\n && \mathcal{L} - \rho\left(\frac{|\u |^{2}}{2} - e - \phi - \frac{p}{\rho} - \lambda \thermod{p}{\rho}{s}\  \right). 
\label{eq: constrained generalized pressure}
\Eeq
Simplifying \eq{eq: constrained generalized pressure} and using \eq{pi-def} we find,
\Beq
\label{GPI-pressure} \mathcal{P} = p + \tilde{p} = p_{0} + \tilde{p}
\Eeq
Comparing to \eq{Euler pressure}, \eq{GPI-pressure} suggests that $\tilde{p}$ indeed carries the closest possible correspondence to the pressure fluctuation $\tilde{p}   \leftrightarrow p^{\prime} \equiv p - p_{0}$ in the fully compressible dynamics. We show in the next section that the correspondence between the Lagrange multiplier and pressure perturbations arises naturally from considering enthalpy in the Lagrangian framework.

For the GPI equations, the potential vorticity becomes,
\Beq
\label{GPI-PV}
q  = \frac{c_p}{\rho T\alpha_T}
 \left(\frac{\grad p_0}{\Gamma_1 p_0}-\frac{\grad\rho}{\rho}\right)\dot( \curl\u),
\Eeq
where $c_{p}$,  $T$, $\alpha_{T}$, and $\Gamma_1$ all implicitly assume $p=p_{0}$.  Given \eq{GPI-PV}, \eqss{eq:new-momentum}{eq:new-pressure} imply $\partial_{t} q + \u \dot \grad{ q} = 0$ directly from the underlying Lagrangian structure, \ie \eq{GPV conservation}.

\subsection{Enthalpy}
\label{sec: Enthalpy}

In \sec{Lagrangian Derivation of the Euler Equations} we express the Lagrangian density in terms of the fluid's internal energy, and hence its density and entropy.  Here, we provide an alternative derivation in terms of enthalpy. \citet{Salmon&Smith_1994} used the same transformation in the Hamiltonian derivation a nonhydrostatic pressure-coordinate model.  This requires us to introduce the pressure as a canonical    variable, in addition to $\vxi$, and requires stationarity of the action with respect to this additional degree of freedom.  This approach contains a number of advantages in helping to clarify the meaning of the Lagrange multiplier, and also points the way to more sophisticated approximation schemes in the future.

Therefore, we make the following replacement 
\Beq
e(\rho,s) \to h(p,s) - \frac{p}{\rho}
\Eeq
in \eq{L-density-0} to produce
\Beq
\label{enthalpy-L} \mathcal{L}(\u,\rho,s,p) = \rho \frac{|\u |^{2}}{2} - \rho h(p,s) - \rho \phi + p.
\Eeq
Stationarity of the action with respect to $p$ exactly enforces the equation of state given in \eq{rho(p,s)-eq},
\Beq
\label{Lagrangian pressure constraint}\pd{\mathcal{L}}{p} = -\rho \thermod{h(p,s)}{p}{s} + 1 = 0.
\Eeq
In this sense, the pressure acts as a nonlinear Lagrange multiplier; similar to a generalization of $\lambda\, \C(\rho,s,\vec{x})$ in \sec{Lagrangian Derivation of the Pseudo- Incompressible Equations}.  

We make the pseudo-incompressible assumption   
\Beq
p \to p_{0}(\vec{x}) + p_{1}, \quad \mathrm{with}\quad
|p_{1}| \ll |p_{0}|.
\Eeq
Expanding the Lagrangian density in \eq{enthalpy-L} to linear order in $p_{1}$ gives
\Beq
 \mathcal{L} \approx \rho \frac{|\u |^{2}}{2}\! -\! \rho (h(p_{0},s)  + \phi)\!-\! p_{1}\!\left[ \frac{\rho}{\rho(p_{0},s)} \! - \! 1  \right] \! + \!p_{0}.\quad\ \ \label{L approx}  
\Eeq
Now $p_{1}$ acts as the Lagrange multiplier enforcing the constraint  
\Beq
\label{rho-p0-s} \inv{\rho(p_0,s)} = \thermod{h}{p}{s}\,\bigg|_{p=p_{0}} = \inv{\rho}.
\Eeq
\Eq{rho-p0-s} requires some interpretation.  The density on the right-hand side of \eq{rho-p0-s}  (representing mass per unit volume) remains frozen into the flow, and evolves according to \eq{rho-eq}.  \citet{Durran_1989} used the notation $\rho^{*}$, to distinguish this from the density in the equation of state.  The density on the left-hand side of \eq{rho-p0-s} represents the equation of state density derived from enthalpy.  \Eq{rho-p0-s} implies that the dynamical mass density must instantaneously equilibrate to the potential density of a parcel adiabatically transported to a reference pressure level, $p_{0}$.  This is mathematically equivalent to the constraint $p(\rho,s) = p_{0}(\vec{x})$ imposed in \sec{Lagrangian Derivation of the Pseudo- Incompressible Equations}.

Furthermore, substituting \eq{L approx} into the general Euler--Lagrange \eq{Euler--Lagrange} gives exactly the same result as \eq{new-u-eq} only with $p_{1}$ replacing $\tilde{p}$.  This underlies the correspondence between the Lagrange multiplier and the small dynamic pressure fluctuations.  Understanding this correspondence becomes necessary when one wants to import data from a sound-filtered numerical model into a fully compressible simulation.  The fully compressible simulation will generate significant acoustic transients if the initial conditions start out of pressure balance.  Using \eq{new-u-eq} with the extra momentum source relative to the LM \eq{eq:LM momentum}, allows the pressure field to donate just enough momentum to the flow to keep it from oscillating rapidly if it were allowed.  

In principle, one could choose to carry the expansion of enthalpy around the background pressure to higher-order in $p_{1}$.   This would reintroduce linear acoustic dynamics back into the equations of motion.  This gives an intermediate approach between the FC equations and the GPI equations, and in principle, would allow a scale-by-scale filtering of acoustic modes according to some criterion such as their timescale.  We leave these types of considerations for future work. 

\subsection{Lagrangian Derivation of the Anelastic Equations}
\label{sec: Lagrangian Derivation of the Anelastic Equations}

We next tie our current results back to Part~I, showing  how the LBR anelastic equations naturally follow from an additional linearization around the background entropy profile.  

We point out that the constraint in \eq{L approx} almost constrains the density to give the background value, except for the entropy dependence.  Therefore, if we make the replacement in the constraint term,
\Beq
\label{constraint replacement} p_{1} \left[ \frac{\rho}{\rho(p_{0},s)} - 1 \right] \to p_{1} \left[ \frac{\rho}{\rho(p_{0},s_{0})} - 1 \right], 
\Eeq
then stationarity with respect to $p_{1}$ would yield $\rho = \rho_{0}$ and $\div ( \rho_{0}\u) = 0$.  Making the replacement in \eq{constraint replacement} produces a version of the LBR anelastic equations with a nonlinear buoyancy term deriving from enthalpy term in \eq{L approx}.
However, we cannot justify \eq{constraint replacement} without admitting that 
\Beq
\label{s-approx} \rho \approx \rho_{0}  \implies  s \approx s_{0}
\Eeq
in some fashion.  In particular, we must guarantee that \eq{s-approx} is consist with the entropy equation.

In Lagrangian coordinates, the dynamics conserves the specific entropy of a fluid parcel along a given trajectory.  That is, 
\Beq
s(t,\vec{x}(t,\vec{a})) = s_{0}(\vec{a}),
\Eeq
where $\vec{a} = \vec{x}(t=0,\vec{a})$ represents the initial position of a fluid parcel.
Therefore, defining an entropy perturbation 
\Beq
s^{\prime}(t,\vec{x}(t,\vec{a})) \equiv s_{0}(\vec{a}) - s_{0}(\vec{x}(t,\vec{a})),
\Eeq
the mean-value theorem implies that 
\Beq
\label{small ds} |s^{\prime}| \le L_{z} \max_{\vec{x}} |\grad{s_{0}}(\vec{x})|,
\Eeq
if we assume the background profile only depends on the height within of the layer.  Here $L_{z}$ represents the total layer thickness.

Therefore, if we assume a small background entropy gradient, then we may further linearize the Lagrangian density in \eq{L approx} and obtain 
\Beq
\label{L anelastic} \mathcal{L} \approx \rho_{0} \frac{|\u |^{2}}{2} - \rho_{0} T_{0} s  - p_{1}\left[ \frac{\rho}{\rho_{0}}  - 1  \right] +\mathcal{L}_{0}\quad,  
\Eeq
where $\mathcal{L}_{0} = -\rho_{0}( e_{0} - T_{0} s_{0} + \phi) $ gives a dynamically irrelevant offset.  
\Eq{L anelastic} implies that we neglect energy sources from nonlinear terms in the thermodynamic variables.
Extremizing \eq{L anelastic} produces the compact form of the LBR formulation of the anelastic equations, 
\Beq
\label{anelastic-u} &&\frac{D \u }{Dt} + \grad{\varpi} \ = \ - s^{\prime}\, \grad{T_{0}}\n
\label{anelastic-s} && \frac{D s^{\prime}}{Dt} + \u \dot \grad{s_{0}} \ = \ 0 \n 
\label{anelastic-rho} && \div{\left(\rho_{0} \u \right)} \ = \ 0,
\Eeq
where often 
\Beq
\label{varpi-eq} \varpi \equiv \frac{p_{1}}{\rho_{0}}
\Eeq
denotes a kinematic pressure variable, and $\rho_{1} = p_{1}/c_{0}^{2}$ gives the physical density perturbations associated with the evolution in \eqss{anelastic-u}{varpi-eq}.   This derivation remains valid only if $s'$ remains small.  By \eq{small ds}, this can occur in general only if $\grad s_0$ is small compared to $L_{z}^{-1}$.  This holds clearly in a nearly adiabatic atmosphere, where $\grad s_0$ is tiny.  Alternatively, the assumption of small $s^{\prime}$ may hold (weakly) for motions with small vertical displacements compared to $L_z$.

For an ideal gas,
\Beq
\grad{T_{0}} = - \frac{\grad{\phi}}{c_{p}} = \frac{\g}{c_{p}}
\Eeq
gives the temperature gradient in the more common form for an adiabatic background atmosphere.
This then puts the momentum equation in a more familiar form
\Beq
\frac{D \u }{Dt} + \grad \varpi \ = \ - \frac{s^{\prime} \g}{c_{p}}. 
\Eeq
For a radiation-pressure-dominated gas, $c_{p} \to \infty$, however 
\Beq
\grad T_{0} = \frac{\g}{s_{0}},
\Eeq
for the nearly constant background entropy $s_{0}$. As for the LBR equations of Part~I, the LBR \Eqss{anelastic-u}{anelastic-rho} also conserve the energy 
\Beq
E \ = \ \rho_{0}\left(\frac{|\u |^{2}}{2}  +  s^{\prime}T_{0} \right),  
\Eeq
and an associated free energy.  Whilst the buoyancy term in \eq{anelastic-u} may appear surprising on first glance, temperature represents the only physically meaningful quantity for nearly adiabatic stratifications that produces an energy when multiplied with entropy fluctuations.

When compared to Part~I, \eq{anelastic-u} implies  a severe contradiction when attempting to model an isothermal atmosphere with a ``nearly adiabatic'' background.  This abuse of approximation becomes clear when one notices the complete vanishing of all buoyancy forcing in \eq{anelastic-u} with constant $T_{0}$.  This implies that one may derive the isothermal LBR equations from a Lagrangian that replaces $T_{0} \to - \phi/c_{p}$ in \eq{L anelastic}, but this replacement would not follow from any systematic approximation of the fully compressible dynamics, and would only apply for an ideal gas.  Nevertheless, \eqss{eq:new-momentum}{eq:new-pressure} reduce to the anelastic \eqss{anelastic-u}{anelastic-rho} for a nearly adiabatic background.

\subsection{Boussinesq Equations}
\label{sec: Boussinesq Equations}

At the outset, we only require a Lagrangian density pertaining to a fluid media, which depends on $\rho$ and $s$, but not directly on the Lagrangian displacement as in the case of a solid.  Other than the fluid requirement, we say nothing more specific about the equation of state --- in particular we did not distinguish between liquids or gasses.  The difference between liquids and gasses mainly derives from the strong independence of density as a function of pressure at fixed entropy, or $\Gamma_{1} \gg 1$.  

For example, in case of water under laboratory conditions, 
$p_{0} \approx 10^{5}\, \mathrm{N}/\mathrm{m}^{2}$, $\rho_{0} \approx 10^{3}\,\mathrm{kg}/\mathrm{m^{3}}$,  $c_{0} \approx 1500\, \mathrm{m}/\mathrm{s}$, and therefore $\Gamma_{1} \approx 2.25 \times 10^{4}$.  $H_{p} = c_{0}^{2}/g \approx 225\,\mathrm{km}$,
However, over the temperature range of liquid water, the density can very by up to $10\%$,  with a density maximum at $T \approx 4^{\mathrm{o}}\mathrm{C}$ allowing for penetrative convection\citep{Veronis_1963}.  Therefore, nontrivial density backgrounds may exist in modelling liquids, but such systems may still filter acoustic modes and may not undergo any significant volume changes.

We therefore note that the GPI equations directly imply generalized Boussinesq equations in the limit  
\Beq
\Gamma_{1} \to \infty.
\Eeq
Dropping all terms involving $\Gamma_{1}$ in \eqss{eq:new-momentum}{eq:new-pressure} gives
\Beq
\label{eq:boussinesq-momentum} && \rho \frac{D\u }{Dt} + \left(\rho - \rho_{0}\right) \grad{\phi} = - \grad{\tilde{p}}, \n
\label{eq:boussinesq-density} &&\frac{ D \rho}{Dt}  =  0, \n
\label{eq:boussinesq-pressure} && \div{\u}  = 0,
\Eeq
Unlike the standard Boussinesq model, $\rho_{0}$ may depend nontrivially on $z$, and large density fluctuation may occur, especially in the presence of combined compositional and thermal effects.

\section{Linear Waves in isothermal atmospheres}
\label{sec: isothermal atmosphere}

As in Part~I, we illustrate the properties of various systems of equations with the dynamics of gravity waves in bounded plane-parallel atmospheres assuming an isothermally stratified ideal gas.  These simple atmospheres possess  constant sound speed, buoyancy frequency, and density scale height.  This makes computing eigenfrequencies and eigenmodes for linear gravity and acoustic waves (when present) analytically tractable. This helps elucidate the differences between the various soundproof equations. These analytic results set the stage for the numerical experiments of \sec{bounded atmospheres and numerics}. We use an ideal gas equation of state
\Beq
p=(c_{p}-c_{v}) \rho T =  (\gamma-1)\rho e 
\Eeq
with specific internal energy $e$, and $\gamma=c_{p}/c_{v}=5/3$.  

As in Part~I, we define the velocity in terms of the vector displacement,
\Beq
  \u = \pd{\vxi}{t}, 
  \label{eq:xi}
\Eeq
which allows the simple integration of the linear density and pressure equations.

We assume wavelike perturbations 
\Beq
\vxi \propto \Xi(z) \exp{\left(i\omega t  - i m x\right)},
\label{eq:wave properties}
\Eeq
where $x$ represents the horizontal coordinate, and $m$ its associated wavenumber, and $\Xi(z)$ gives the vertical structure depending on the specific model.  

In a hydrostatically balanced (\eq{eq: hydrostatic}) isothermal atmosphere 
\Beq
  \grad{ \ln p_{0} } = \grad{ \ln \rho_{0} } = 
  -\frac{\vec{\hat{z}}}{H} =  -\frac{\gamma g\, \vec{\hat{z}}}{c_{0}^2}
  \label{eq:isothermal scaleheight}
\Eeq
where $H$ is the (constant) pressure or density scale height,  and
\Beq
  c_0^2 \equiv \frac{\gamma p_{0}}{\rho_0} 
\Eeq
is the (constant) sound speed.
The Brunt-V\"ais\"al\"a frequency $N_{0}$ is
\Beq
  N_{0}^2 = -  \frac{\g \dot\grad{s_0}}{c_{p}} =
  \frac{(\gamma -1)}{\gamma} \frac{g}{H}
\label{eq:N2}
\Eeq
where $\grad{ s_{0} }$ is the background entropy gradient.

\subsection{Fully compressible waves}
\label{sec: Fully compressible equations}
 
The well-known solution for the fully compressible equations exists in several textbooks
\citep[e.g.,][]{Lighthill_1978_Waves_in_Fluids} and in Part~I.  In a bounded or unbounded atmosphere, vertical eigenfunctions of the take the following form
\Beq
\label{vertical e-function}  \Xi(z) =  \exp{\left(\frac{z}{2H}\right)} \sin\!{\left(k z + \delta \right)} ,
\label{eq:compressible bounded vertical eigenfunction}
\Eeq
where $k$ and $\delta$ give an arbitrary vertical wavenumber and phase.  For finite domains with impenetrable boundaries, only quantized modes
\Beq
  k =  \frac{\pi n }{L_{z}},~n=1,2,\ldots, \quad \delta =0,
  \label{eq:k}
\Eeq 
ensure that vertical motions cease ($\xi_z =0$) at $z=0, L_{z}$.  \Eq{vertical e-function} implies that the frequencies of gravity and acoustic modes remain purely real, \ie 
\Beq
-\frac{\omega^4}{c_{0}^2}+\omega^{2}\left[k^{2}+m^{2}+\inv{4H^{2}}\right]=m^{2}N_{0}^{2}.
\label{eq:compressible dispersion}
\Eeq
The quadratic nature of \eq{eq:compressible dispersion} in $\omega^2$ provides the two distinct  acoustic and gravity wave branches.  We write the full solution to both branches of \eq{eq:compressible dispersion} in the form,
\Beq
\omega_{\pm}^2 = \frac{\omega^2_{c}}{2} \left( 1\pm \sqrt{1-\frac{4 \omega^2_{g}}{\omega^2_{c}}}\right),
\label{eq:omega Navier--Stokes infinite isothermal full dispersion}
\Eeq
where
\Beq
  \omega_{c}^2 = \left[k^2 + m^2 +\inv{4H^2}\right]  c_{0}^2, \label{eq:compressible dispersion SW}
\Eeq
and
\Beq
  \omega_{g}^2 = \frac{m^2}{k^2 + m^2 +\inv{4H^2}} N_{0}^2.
  \label{eq:compressible dispersion GW}
\Eeq
\Eqs{eq:compressible dispersion SW}{eq:compressible dispersion GW} represent simple approximations to the behavior of \eqs{eq:compressible dispersion}{eq:omega Navier--Stokes infinite isothermal full dispersion} in the high- and low-frequency limits respectively.  

Importantly, the linear waves conserve the leading-order quadratic energy in \eqss{free energy}{linear-APE}.  

\subsection{Pseudo-incompressible gravity waves}
\label{sec: PI waves}

We begin our linear analysis of the sound-filtered systems with the energy-conserving PI equations.  As we show in \sec{Lagrangian Derivation of the Euler Equations},
the GPI \eqss{eq:new-momentum}{eq:new-density} reduce to the PI \eqss{eq:PI momentum}{eq:PI constraint} for $\Gamma_{1} = \gamma$.
Linearizing the pressure, momentum and
density equations around the background state gives 
\Beq
-\rho_0\omega^2 \vxi &=& -\beta_{0}\grad\left(\frac{p^{\prime}}{\beta_{0}}\right) + \rho_1\g ,
\label{eq:PI wave momentum}
 \n
  \rho_1/\rho_0 &=& -\vxi\dot \grad{ \ln \rho_0} -  \div{\vxi}
 \label{eq:PI wave density}\n
\div{\vxi} &=& -\vxi \dot \grad{\ln \beta_{0}} ,
\label{eq:PI wave continuity}
\Eeq
where $\beta_{0}\equiv p_{0}^{1/\gamma}$ (\eq{beta-def}).
We combine the density \eq{eq:PI wave density} and pressure \eq{eq:PI wave continuity} to give
\Beq
  \frac{\rho_1}{\rho_0} = \vxi \dot \grad \ln \frac{\beta_{0}}{\rho_0} = \vxi\dot \frac{\grad s_{0}}{c_{p}} .
  \label{eq:PI wave buoyancy}
\Eeq
Thus the buoyancy term in the pseudo-incompressible equations remains identical to the buoyancy term in the anelastic equations considered in Part~I.

The divergence of the horizontal momentum equation, and the pressure constraint determines the pressure fluctuations in terms of the vertical displacement, 
\Beq
 \delsq p^{\prime} = \rho_0 \omega^2\grad_{\perp}\dot \vxi_{\,\perp} = -\frac{\omega^{2} \rho_{0}}{\beta_{0}}\pd{(\beta_{0} \xi_{z})}{z}.
  \label{eq:PI pi horizontal divergence}
\Eeq

Combining the vertical momentum \eq{eq:PI wave momentum}
and the buoyancy \eq{eq:PI wave buoyancy} for linearized
waves, we obtain 
\Beq
  \omega^2 \delsq\xi_z - \frac{\beta_{0}}{\rho_0}\pd{}{z} \left(\frac{ \delsq p^{\prime} }{ \beta_{0}}\right) = N_{0}^2 \delsq \xi_{z},
  \label{eq:PI vertical momentum omega, first step}
\Eeq
where we use \eq{eq:N2}.  Collapsing the entire system into a single second-order equation for vertical displacement gives
\Beq
  \omega^2\left( \delsq  + \mathcal{D}_{\mathrm{PI}}^{2}\right)\xi_{z} = N_{0}^2 \delsq \xi_{z},
  \label{eq:PI vertical momentum omega}
\Eeq
where,
\Beq
\mathcal{D}_{\mathrm{PI}}^{2}\xi_{z} \equiv \frac{\beta_{0}}{\rho_{0}}\pd{}{z}\left[ \frac{\rho_{0}}{\beta_{0}^{2}}\pd{(\beta_{0}\xi_{z}) }{z}\right],
\Eeq
is a negative-definite self-adjoint operator with respect to the integration weight function $\rho_{0}(z)$. Self-adjointness follows from the identity 
\Beq
\int\!\rho_{0}\, \eta \,\mathcal{D}_{\mathrm{PI}}^{2} \xi \dd{z} = -\! \int\! \frac{\rho_{0}}{\beta_{0}^{2}} \pd{(\beta_{0} \eta)}{z} \pd{(\beta_{0} \xi)}{z}\dd{z},\quad
\Eeq
assuming $\eta = \xi = 0$ on the endpoints of the integration domain.  

For the isothermal background in \sec{Fully compressible equations}, the same vertical eigenfunction profile in \eq{vertical e-function} provides the real-valued dispersion relation
\Beq
  \omega_\mathrm{PI}^2 \left[ m^2 + k^2 + \frac{(\gamma-2)^2}{4\gamma^{2}H^{2}}\right] = m^2 N_{0}^2,\quad \ 
  \label{eq:PI dispersion relationship}
\Eeq
which closely resembles \eq{eq:compressible dispersion GW}, but with a different large-scale cutoff.  We note that the correspondence between PI and FC eigenfunctions breaks down in spherical geometries.  In spherical geometries, the radial structure of the FC eigenfunctions depend explicitly on the spherical harmonic degree, $\ell$, but not in the PI approximation. However, many of the fundamental qualitative properties developed here remain even in those systems.  

In a monatomic ideal gas with $\gamma=5/3$, \eq{eq:PI dispersion relationship} implies a large-scale cutoff frequency $1/(10 H)$; the cutoff vanishes when $\gamma=2$, and approaches $1/(2H)$ for larger $\gamma$.  The vanishing of the large-scale cutoff for $\gamma=2$ actually corresponds to similar behavior in the fully compressible dispersion relationship for any $m$ as $k \to 0$.  More generally, for any background, \eq{eq:PI dispersion relationship} gives an optimal approximation to the FC frequencies in the sense that
\Beq
\label{omega-error} | \omega_{-} - \omega_{\mathrm{PI}}| = \mathcal{O}\left(m^{-4}\right), \quad \mathrm{as}\quad m  \to \infty.
\Eeq
More generically, the error in the difference between the approximate and fully compressible frequencies scales as $\sim m^{-2}$. 

\begin{figure}
  \begin{center}
    \includegraphics[width=9cm]{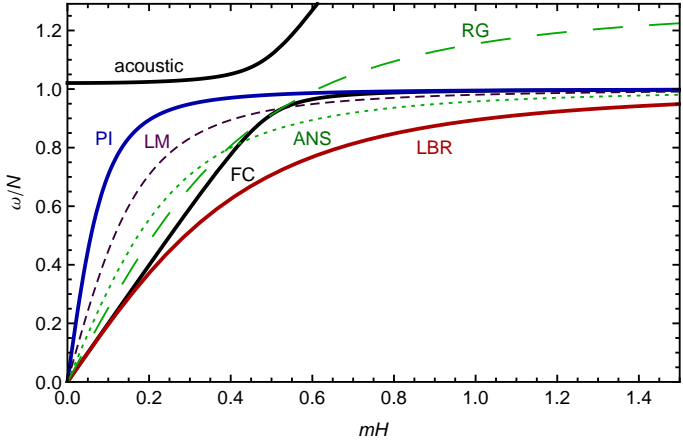}
  \end{center}
  \caption{
  Dispersion relationships for waves with $kH \to 0$ for all equation sets considered here and in Part~I
  (and see Table~\ref{table:dispersion relationships}).  
  Here we also show the sound wave branch (black, labelled ``acoustic'') of the exact solution to the full Euler equations.
\label{fig:all equations omega infinite atm}}
\end{figure}

\fig{fig:all equations omega infinite atm}{} shows the dispersion relationship for the PI \eq{eq:PI dispersion relationship}, in the limit $k H \to 0$, and compares this to the equivalent modes in FC, LM, and all other anelastic models considered in Part~I.  \fig{fig:all equations omega infinite atm}{} clearly shows the behavior in \eq{omega-error} relative to other models for large $m H$.
At low horizontal wave number $m$, the reduced large-sale cutoff in the PI dispersion relationship (\eq{eq:PI dispersion relationship}) leads to higher
frequencies than those obtained from the FC 
\eq{eq:omega Navier--Stokes infinite isothermal full dispersion}.
For particular comparison, we also plot the dispersion relationship of the
energy-conserving LBR anelastic equations from Part~I. Generally,
frequencies obtained from the LBR equations match the Euler
frequencies well at low frequencies but approach $\omega \sim N_{0}$ slowly for large $m$. The differences in frequencies obtained in all energy-conserving equation sets become smaller as $k H$ increases. We also show all of the equation sets considered in this paper and in Part~I in 
Figure~\ref{fig:all equations omega infinite atm} for the same isothermal atmosphere.  
The PI, FC and LBR equations conserve energy, while the LM and ANS equations do not.  
In general, the RG equations do not conserve energy, except in the special case of an isothermal atmosphere (see Part~I).  At large $m H$, the ANS, LBR, LM, PI and FC equations converge to the Brunt-V\"ais\"al\"a frequency $N_{0}$, while the RG equations are too large by a factor of $\sqrt{\gamma}$. 
Table~\ref{table:dispersion relationships} gives the dispersion relationships for all equation sets in both papers.

\begin{figure}
  \begin{center}
    \includegraphics[width=9cm]{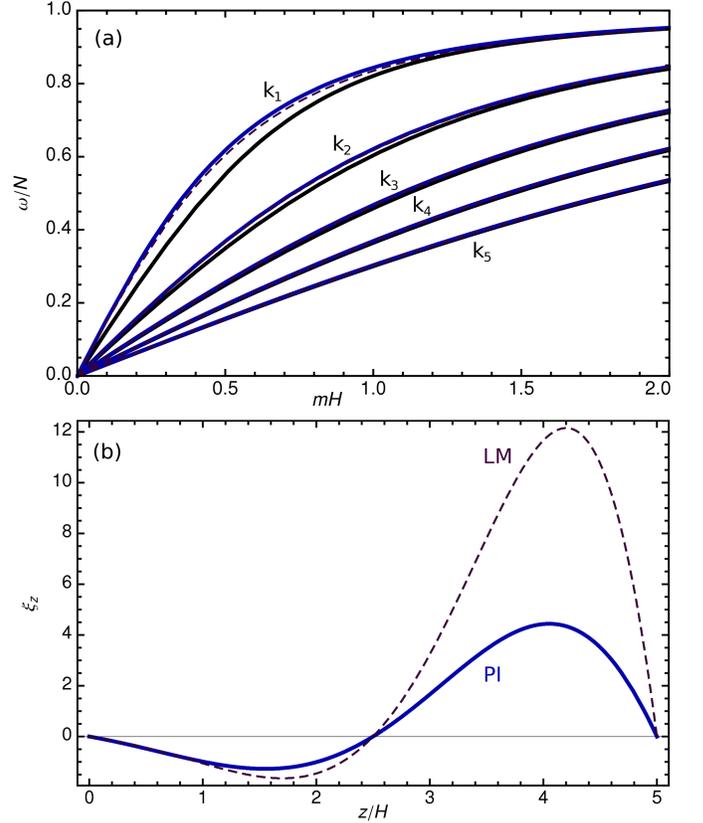}
  \end{center}
\caption{Properties of waves in bounded atmospheres.  (a) Dispersion
  relationships for PI equations (blue, solid), LM equations (purple,
  dashed) and full Euler FC equations (black, solid) for waves with vertical wavenumbers $k_1$--$k_5$, as labelled, plotted against
  scaled horizontal wavenumber $mH$.  The numerical simulations of
  \sec{bounded atmospheres and numerics} use $k_2$ and
  $mH\approx 1.26$.  (b) Vertical eigenfunctions for the PI and LM equations for the $k_2$ mode with   \eq{eq:eigenfunction normalization} giving the normalization.  Eigenfunctions of the FC equations
  are identical to the PI equations and are not shown.  Though the frequencies of the LM equations are reasonably similar to those obtained in the energy conserving PI and FC equations, the LM eigenfunctions differ significantly.  \label{fig:bounded atmosphere}
}
\end{figure}

\fig{fig:bounded atmosphere}{a} shows the first five vertical modes $k_1$--$k_5$ of the dispersion relationship for the PI, LM and FC equations, where \eq{eq:k} gives $k$ with $L_{z} = 5 H$. 
Similar to \fig{fig:all equations omega infinite atm}, the frequencies of gravity waves in the PI equations remain somewhat larger than the FC equations at low horizontal wave number, with the discrepancy most notable for long-vertical-wavelength modes (e.g., $k_1$).  \fig{fig:bounded atmosphere}{b} shows the vertical eigenfunction structure for the $k_2$ mode.  This mode, and the stratification of this atmosphere, 
correspond to the numerical results we present in
\sec{bounded atmospheres and numerics}.  

As in Part~I, we normalize these eigenfunctions with an amplitude $A$, such that
\Beq
  A^2 = \frac{\int_0^{5H}{\Xi{(z)}^2 e^{- \epsilon z/H}dz}}{\int_0^{5H}{e^{-\epsilon z/H}dz}}
  = \frac{5}{2} \frac{\epsilon}{1-e^{-5 \epsilon}}.
  \label{eq:eigenfunction normalization}
\Eeq
In the PI and FC equations, $\epsilon=1$, and $A_{\mathrm{PI}} \approx 1.59$, for $\gamma=5/3$
whereas in the LM equations $\epsilon=(\gamma+1)/\gamma$, and $A_{\mathrm{LM}}\approx 2.00$. 

\subsection{Low Mach number gravity waves}
\label{sec: LM waves}

Finding linear eigenfrequencies in the LM \eqss{eq:LM momentum}{eq:LM divergence} amounts to the
same procedure as in \sec{PI waves}.  
Now however the linearized momentum equation is
\Beq
-\rho_0 \omega^2 \vxi = -\grad p^{\prime} +\rho_1\g .
   \label{eq:LM momentum omega}
\Eeq 

The fundamental differences in the pressure term 
between \eqs{eq:PI wave momentum}{eq:LM momentum omega} 
lead to the non-conservation of energy in the LM equations and instead to the conservation of pseudo-energy for linear waves. 

Using the linearized \eqs{eq:PI wave density}{eq:PI wave continuity} for density and pressure, and \eq{eq:LM momentum omega} for momentum, we collapse the LM system into a form equivalent to \eq{eq:PI vertical momentum omega},  
\Beq
  \omega^2\left( \delsq  + \mathcal{D}_{\mathrm{LM}}^{2}\right)\xi_{z} = N_{0}^2 \delsq \xi_{z},
  \label{eq:LM vertical momentum omega}
\Eeq
where,
\Beq
\mathcal{D}_{\mathrm{LM}}^{2}\xi_{z} \equiv \inv{\rho_{0}}\pd{}{z}\left[ \frac{\rho_{0}}{\beta_{0}}\pd{(\beta_{0}\xi_{z}) }{z}\right],
\Eeq
is a negative-definite self-adjoint operator with respect to the modified integration weight function,
\Beq
\label{rho-0-hat} \hat{\rho}_{0}(z) \equiv \rho_{0}(z) \beta_{0}(z).
\Eeq
That is, assuming $\eta = \xi = 0$ on the endpoints of the integration domain,
\Beq
\int\!\hat{\rho}_{0}\, \eta \,\mathcal{D}_{\mathrm{LM}}^{2} \xi \dd{z} = -\! \int\! \frac{\hat{\rho}_{0}}{\beta_{0}^{2}} \pd{(\beta_{0} \eta)}{z} \pd{(\beta_{0} \xi)}{z}\dd{z}.\quad \quad
\Eeq

For an adiabatic background atmosphere, $\hat{\rho}_{0} \propto \rho_{0}^{2}$ and the LM equations do not reduce to an anelastic system.  This implies that the LM equations would not give correct growth rates and vertical structure for convection in an unstably stratified system.

For an ideal gas, \eqss{eq:LM vertical momentum omega}{rho-0-hat} require the vertical eigenfunction 
\Beq
  \Xi(z) = 
  \exp{\left(\frac{\gamma+1}{\gamma}\frac{z}{2H}\right)}\sin\!\left(k z + \delta \right),
 \label{eq:LM bounded vertical eigenfunction real omega}
\Eeq
to guarantee real-valued wave frequencies, 
\Beq
   \omega_\mathrm{LM}^2\left[m^2+ k^2 + \frac{(\gamma-1)^{2}}{4 \gamma^{2}H^2}\right] = m^2 N_{0}^2.
   \label{eq:LM dispersion relationship real omega}
\Eeq

As with the ANS equations in Part~I, a serious problem lurks with this choice, since the kinetic energy density scales as 
\Beq
  \rho_0 u^2 \propto \exp\left(\frac{z}{\gamma H}\right).
\Eeq
Therefore, the kinetic energy of the waves grows exponentially with height.  From the FC equations, we know that this kinetic energy density should remain constant with height in an isothermal atmosphere.  This discrepancy arises from the fact that the LM equations fail to properly conserve energy, but rather conserve the pseudo-energy given in \eq{pseudo-energy}.

\fig{fig:bounded atmosphere}{a} shows 
the dispersion relationship with real $\omega$ given in \eq{eq:LM dispersion relationship real omega} for an isothermal atmosphere with $L_{z}=5H$. Like the PI equations, the LM equations give slightly higher frequencies than the fully compressible equations.  A much larger problem however rests with the eigenfunctions of the waves. \fig{fig:bounded atmosphere}{b} shows the eigenfunctions in an isothermal atmosphere with $L_{z}=5H$.  \Eq{eq:eigenfunction normalization} provides the normalization, and gives the response of waves in the LM equations generated from the same initial density perturbation used for the PI equations. Clearly, the eigenfunctions of waves in the LM equations acquire an excess amplitude  compared to the correct eigenfunctions given in \eq{eq:compressible bounded vertical eigenfunction}.

\begin{deluxetable}{ccccccccccccc}
\tabletypesize{\footnotesize}
 \tablecolumns{3}
 \tablewidth{0pt}  
 \tablecaption{dispersion relationship comparison
 \label{table:dispersion relationships}}
 \tablehead{\colhead{System}  &  
   \colhead{$ (\omega / m N_{0})^{-2} =$} &
   \colhead{eq} 
}
\startdata
FC &
see text
& (\ref{eq:omega Navier--Stokes infinite isothermal full dispersion}) \\[2mm]
PI & 
$m^2 + k^2 + \frac{(\gamma-2)^{2}}{4 \gamma^{2} H^2}$
& (\ref{eq:PI dispersion relationship}) \\[2mm]
LM &
$m^2+ k^2 + \frac{(\gamma-1)^{2}}{4 \gamma^{2} H^2}$
&  (\ref{eq:LM dispersion relationship real omega})\\[2mm]
ANS &
 $m^2+ k^2 +\inv{4 \gamma^2 H^2}$
& (P1: 56) \\[2mm]
LBR & 
$m^2 + k^2 + \inv{4 H^2}$
& (P1: 62) \\ [2mm]
RG & 
$\inv{\gamma}\left(m^2 + k^2 + \inv{4 H^2}\right)$
& (P1: 67) \\
\enddata
\tablecomments{We quote the ANS and LBR dispersion relationships from Part~I (P1). \\
}
\end{deluxetable}

\section{Simulations of gravity waves in a bounded isothermal atmosphere}
\label{sec: numerical}
\label{sec: bounded atmospheres and numerics}

In this section we demonstrate through direct numerical simulation the
properties of the PI and LM equations described in \sec{Pseudo-incompressible equations}, \ref{sec: MAESTRO low Mach number equations}, \& \ref{sec: isothermal atmosphere}.
We use the MAESTRO code to simulate the propagation of internal
gravity waves in a bounded isothermal atmosphere, implementing both
the PI and LM equations (see Appendix~\ref{sec: modifications} for more details).  Although MAESTRO supports adaptive mesh
refinement (AMR) and an evolving base state, we use neither of these
capabilities here \citep{Almgren_et_al_2008, Nonaka_et_al_2010, Nonaka_et_al_2012}.  In our implementation of the PI equations , we take an ideal gas equation of state, assume that $\Gamma_1 (p_0,\rho)=\gamma = 5/3$, and solve the nonlinear PI \eqss{eq:PI momentum}{eq:PI constraint} and LM \eqss{eq:LM momentum}{eq:LM divergence}.
In both cases we do not include any explicit diffusivities, and rely instead on numerical diffusivities within the MAESTRO code itself.

We use an isothermal, hydrostatically balanced background state with constant scale height, $H$, and gravitational acceleration, $g$, such that
\Beq
\label{numerical-density-pressure} \rho_{0}(z) = \rho_{\mathrm{ref}} \exp(-z/H),\quad  p_{0}(z) = g H \rho_{0}(z).
\Eeq
Although the PI and LM equations allow arbitrarily large density perturbations, both assume small pressure perturbations, and thus $\beta_{0}(z) \propto \exp(-z/\gamma H) $ remains fixed throughout our simulations.
In this paper we focus on a set of 2D simulations in a Cartesian domain with size $L_{x} \times L_{z}$, where $L_{x} = L_{z} = 5 H$.

For any given background variable, say $\rho$, defining the number of scale heights of this variable proves helpful in discussing the degree of stratification in our simulations, 
\Beq
n_{\rho} \equiv \ln \frac{\rho_{0}|_{z=0}}{\rho_{0}|_{z=L_{z}}}.
\Eeq
Therefore, $n_{\rho} = n_{p} = 5$ in this isothermal atmosphere.  By comparison, for the radiative zone of the Sun, $n_{\rho} \approx 6.6$. 

In the LM equations, $\beta_0$ scales the pseudo-density according to \eq{rho-0-hat}.  The number of pseudo-density scale heights $n_{\hat{\rho}}$ is
\Beq
n_{\hat{\rho}} = n_\rho + n_\beta = n_\rho+\inv{\gamma}n_p,
\Eeq
where $n_\beta$ and $n_p$ represent the number of $\beta$ and pressure scale heights respectively.  When $n_\rho=5$, $n_\beta=3$, and $n_{\hat{\rho}}=8$.  Thus, the gravity waves in the LM simulations experience the effects of stratification much more strongly than the
gravity waves in the PI simulations. 

We conduct the simulations in a Cartesian box with resolution $(N_x,N_z)=(512,512)$.  The box has a periodic horizontal direction, with impenetrable vertical boundary conditions at the top and bottom.  
Since we do not include explicit viscosity or thermal diffusion, we do not require any further boundary condition at the lower or upper boundaries.  We initialize the various simulations
with one of three small density perturbations, 
\Beq
\label{rho-0-PI} \delta\rho_{\mathrm{PI}} & = & A_0 \sin\left(\frac{2\pi x}{H}\right) \sin\left(\frac{2 \pi z}{5H}\right)\exp\left[\frac{z}{2H}\right],\\
\label{rho-0-LM} \delta\rho_{\mathrm{LM}} & = & A_0 \sin\left(\frac{2\pi x}{H}\right) \sin\left(\frac{2 \pi z}{5H}\right)\exp\!\left[\!\frac{(\gamma+1)z}{2\gamma H}\!\right]\!,\quad\quad  \\
\label{rho-0-GW} \delta\rho_{\mathrm{GW}} & = & A_0 \sin\left(\frac{2\pi x}{H}\right) \sin\left(\frac{2\pi z}{5H}\right),
\Eeq
where $A_0=10^{-6} \rho_{\mathrm{ref}}$, and $\rho_{\mathrm{ref}}$ gives the density at $z=0$ from \eq{numerical-density-pressure}.  The perturbations
$\delta\rho_{\mathrm{PI}} $ and $\delta\rho_{\mathrm{LM}} $ correspond
to the $n=2$ eigenfunctions derived in \sec{isothermal atmosphere} 
for the PI and LM equations respectively, with an $m=10$ horizontal perturbation.  Importantly, each PI (LM) eigenfunction projects onto many LM (PI) eigenfunctions, so the PI (LM) eigenfunction excites many modes of the LM (PI) equations.  The third perturbation,
$\delta\rho_{\mathrm{GW}}$, projects onto many modes from both models, and thus excites a broad band of modes in either the PI or LM equations.
We normalize the background density to one at the bottom of the domain, \ie $\rho_{\mathrm{ref}}(0) = 1$.
Because the density perturbations produce small velocity amplitudes, the default velocity-based CFL time-step condition insufficiently captures the Brunt-V\"ais\"al\"a timescale of the gravity waves, where $\tau_{BV} = N_{0}^{-1}$.  We thus restrict our time steps such that $\Delta t \le 0.016N_{0}^{-1}$, which accurately resolves the relevant waves.

\begin{figure}
  \begin{center}
    \includegraphics[width=\linewidth]{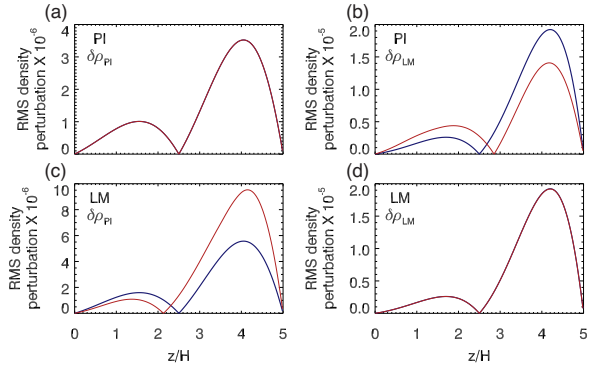}
  \end{center}
\caption{The rms density perturbation at time $t=0$
  (blue) and time $t=600N_{0}^{-1}$ (red).  In $(a,b)$ the simulations
  employ the PI equations, and in $(c,d)$ the simulations employ the LM equations.  Cases $(a,c)$ each use the initial perturbation $\delta\rho_{\mathrm{PI}} $,  $(b,d)$ use $\delta\rho_{\mathrm{LM}}$.  Cases $(a)$ and $(d)$, show no change in form with time, indicating the perturbations coincide with eigenfunction of the respective equations.  In $(b)$ and $(c)$ the initial perturbation differ from eigenfunctions of the relevant equations and so instead excite a band of modes, which interfere at later times.
\label{fig:eigenfunctions}
}
\end{figure}

The analysis in \sec{isothermal atmosphere} shows that the vertical eigenfunctions differ for the two sets of equations. In \fig{fig:eigenfunctions}{} we plot the rms value of the
density perturbation at two times, $t=0$ and $t=600N_{0}^{-1}$, 
in both (PI and LM) types of simulations using initial density 
perturbations based on each eigenfunction in \eqs{rho-0-PI}{rho-0-LM}.  
We compute the rms perturbation by taking the square root of the horizontally averaged
potential energy density (\eq{linear-potential energy}) times $2N_{0}^2/(\rho_0g^2)$.  In each set of equations, when we perturb the
background with the correct eigenfunction, the perturbation remains
the same after an integer number of periods.  We demonstrate this for the PI equations with initial perturbation $\delta\rho_{\mathrm{PI}}$
(Fig.~\ref{fig:eigenfunctions}$(a)$), and the LM equations with initial perturbation
$\delta\rho_{\mathrm{LM}} $ (Fig.~\ref{fig:eigenfunctions}$(d)$).  Otherwise, the
perturbation excites a broad band of modes which interfere, and the
rms amplitude changes in time (Fig.~\ref{fig:eigenfunctions}$(b,c)$).  
This confirms the predicted differences in the eigenfunctions of the PI and LM equations, 
as given in \eqs{eq:compressible bounded vertical eigenfunction}{eq:LM bounded vertical eigenfunction real omega}. 

Another significant difference between the two sets of equations is that the PI equations conserve energy, whereas the LM equations do not.
Although the LM equations violate energy conservation, they still contain a pseudo-energy quadratic invariant in the linear regime (\eq{pseudo-energy}) which differs from the true energy by a factor of $\beta_{0}$ (see discussion in \sec{MAESTRO low Mach number equations}).

\begin{figure*}
  \begin{center}
    \includegraphics[width=\linewidth]{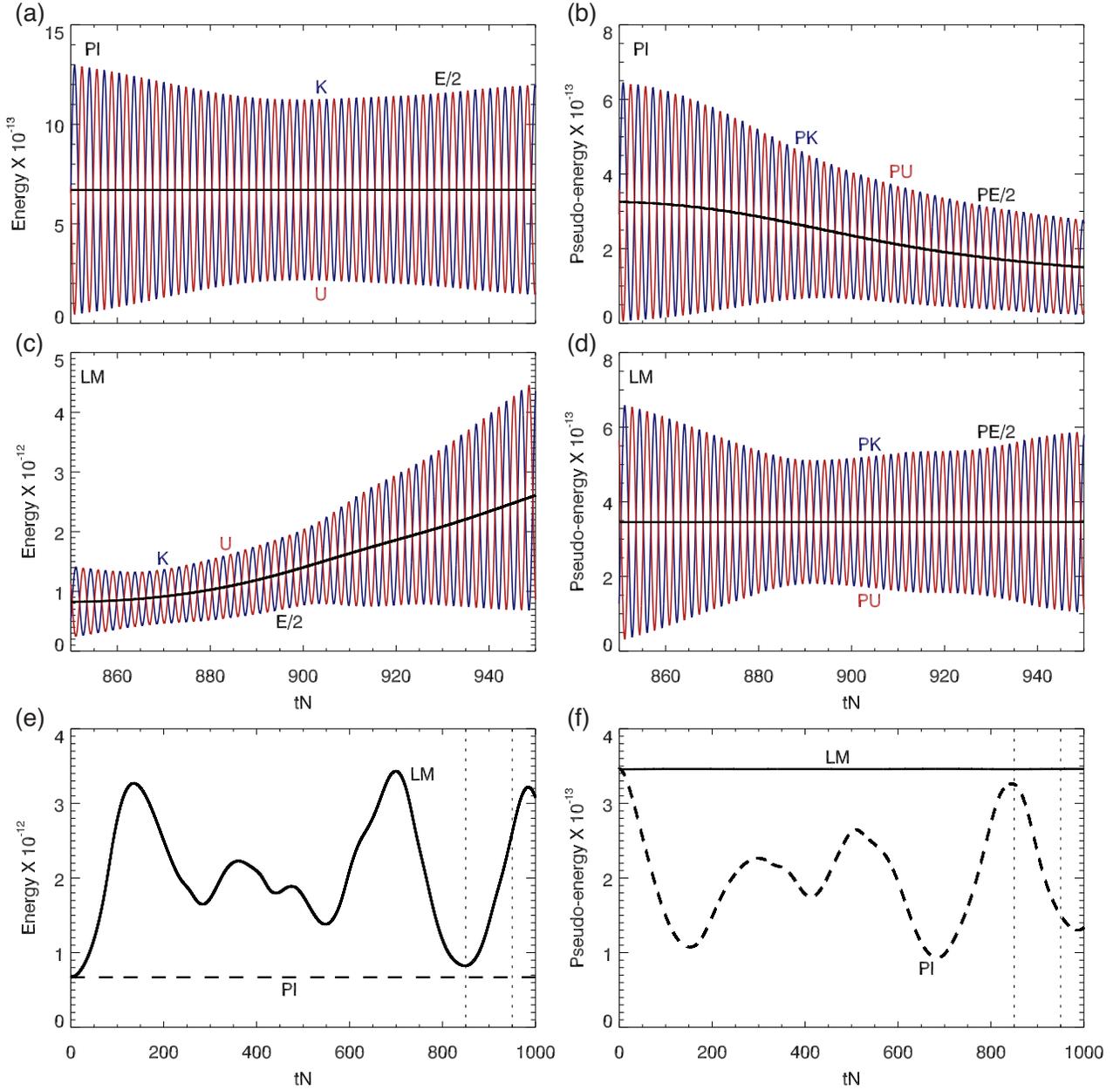}
  \end{center}
\caption{Temporal evolution of energies and pseudo-energies in simulations solving the PI and LM equations with
  $\delta\rho_{\mathrm{GW}}$ as an initial perturbation.  $(a)$ Energy and $(b)$
  pseudo-energy in the simulation solving the PI equations.  The total energy ($E$, \eq{linear-energy}, in black) and pseudo-energy ($\PE$, \eq{pseudo-energy}) are divided by two to highlight the fluctuations between kinetic energy $K$ (\eq{linear-kinetic energy}, in red) and potential energy $U$ (\eq{linear-potential energy}, in blue)
  or $\PK$ (\eq{pseudo-kinetic energy}) and $\PU$ (\eq{pseudo-potential energy}) respectively.  The PI equations correctly conserve total energy, but not the total pseudo-energy.  $(c)$ Energy and $(d)$ pseudo-energy in the simulation solving the LM equations.  The LM equations do not conserve the total energy, and instead conserve the total pseudo-energy.  We plot the total energy $(e)$ divided by two and total pseudo-energy $(f)$ divided by two for the LM simulation (solid line) and the PI simulation (dashed line), over a longer timescale.  The vertical dotted lines show the time range ($850N_{0}^{-1}$, $950N_{0}^{-1}$) depicted in $(a$-$d)$.  Averaging over $1000N_{0}^{-1}$, we have that
  $\Delta E\approx 0.035\%$ for the PI simulation and $\Delta
  \PE \approx 32\%$.  Over this same timescale, $\Delta E\approx 36\%$
  for the LM simulation, but $\Delta \PE \approx 0.035\%$. 
  \label{fig:energy}
}
\end{figure*}

We plot temporal traces of volume-averaged energy and pseudo-energy (\eqs{linear-energy}{pseudo-energy}) for simulations implementing the PI equations and the LM equations in \fig{fig:energy}.  We integrate the equations until $t=1000N_{0}^{-1}$, and use
$\delta\rho_{\mathrm{GW}}$ as an initial perturbation, which excites a
broad band of modes for both simulations.  \fig{fig:energy}{a} demonstrates that the PI equations conserve energy, whereas the LM equations clearly violate energy conservation (\fig{fig:energy}{c,e}).  Instead, the LM equations conserve the pseudo-energy (\fig{fig:energy}{d}),
which the PI equations do not conserve (\fig{fig:energy}{b,f}).  Because both simulations use the the same initial perturbation, $\delta\rho_{\mathrm{GW}}$, the two simulations initially contain the same energy, and the same pseudo-energy (\fig{fig:energy}{e,f}).

\fig{fig:energy}{e,f} show the long-time variation in energy and pseudo-energy.  To quantify the deviations in these quantities, we define the energy and pseudo-energy variation as 
\Beq
\Delta E & \equiv & (\delta E)/\langle E\rangle, \\
\Delta \PE & \equiv & (\delta \PE)/\langle \PE \rangle,
\Eeq
where $\langle\dot\rangle$ denotes a temporal mean and $\delta$
denotes the standard deviation in time for the entire dataset from $0 \le t \le 1000N_{0}^{-1}$.  
In the PI simulation depicted in \fig{fig:energy}, 
$\Delta E\approx 0.035\%$, whereas $\Delta \PE\approx 32\%$.  
Conversely, for the simulation implementing the LM
equations, $\Delta E\approx 36\%$ and $\Delta \PE\approx 0.035\%$.

Slowly accumulating errors within the explicit time-stepping method of MAESTRO leads to the slight deviations from perfect energy conservation in the PI simulation and perfect pseudo-energy conservation in the LM simulation.  The small growth in the ``conserved'' quantities becomes even smaller when we decrease the size of our maximum time step.

We compute a variety of simulations using the LM equations with simulations domains of various sizes, and with different initial perturbations.  As in Part~I, we find that the degree of energy non-conservation, measured using $\Delta E$, scales about linearly with $n_\rho$.  That is, larger $n_\rho$ leads to larger the variations in $E$.  In the simulations shown in \fig{fig:energy}, the initial perturbation contains only one horizontal mode.  Including multiple horizontal modes lowers $\Delta E$, since modes with different frequencies combine incoherently.  This does not mitigate the lack of local energy balance.  In all cases, the pseudo-energy-conserving LM waves tend to have larger amplitudes than the energy-conserving PI waves.

\section{Summary and Conclusions}
\label{sec: conclusions}

In the last few decades, the fluid dynamics and magnetohydrodynamics of stellar and planetary interiors has become amenable to direct numerical simulation. Although the simulations do not capture all the realistic parameter ranges --- they tend to overemphasize the role of diffusive processes, for example --- many approximate treatments such as the mixing length theory of convection, Eddington-Sweet circulation, 
turbulent viscosity, and mean-field dynamo theory can now use numerical solutions of the fluid equations to calibrate useful parameterizations. These advances make it possible to probe many dynamical phenomena, including  penetrative convection, chemical mixing, angular momentum transport, and stellar and planetary dynamos, with unprecedented realism.

Many interesting problems involve the coupling between radiative and convection zones. Penetrative convection gives one example; the degree of overshoot affects dynamo-related phenomena such as magnetic pumping, chemical evolution phenomena such as depletion of light elements, and dynamical phenomena such as the generation of gravity waves. Angular momentum balance generally also tends to couple radiative and convection zones; for example, red giant cores should spin up as they contract while red giant envelopes should spin down, but can angular momentum flow between these two zones?

Because acoustic timescales in stars are typically very short compared to other dynamical timescales, direct numerical simulation of stellar interior dynamics is only practical if acoustic waves can be filtered out. Similar issues arise in the study of planetary atmospheres and accretion disks. Over the years, many ingenious and beautiful treatments have been developed to work around this problem. Two such approximations, the anelastic and pseudo-incompressible models, are valid for adiabatically stratified systems, and thus should be quite accurate in stellar convection zones. 
However, in recent years, interest in modeling
stellar dynamos and stellar internal rotation, circulation, and mixing have also demonstrated the need for models which are valid in stably stratified, radiative zones.  Additionally, modeling double-diffusive buoyancy such as fingering convection and semi-convection requires capturing both significant turbulent and gravity wave dynamics.  

In Part~I we studied the anelastic equations and showed that one particular version (the LBR equations) was particularly versatile because it conserves energy. In this paper we extend our treatment to two consider sets of pseudo-incompressible models (PI and LM).

Since there are many methods for achieving  acoustic filtering, which one works best? It has long been known that the fully compressible Euler equations can be derived from Lagrangian analysis. In \sec{Constrained Models} we showed that approximations such as pseudo-incompressibility can be implemented by a Lagrangian analysis with the approximation scheme enforced through constraints. The resulting equations conserve energy in their ideal form and are thus preferable over nonconservative formulations. As an important  illustration of the technique, we extended the
pseudo-incompressible approximation to systems with a generalized equation of state. This extension will enable treatment of gas-radiation mixtures, partially degenerate systems, and other thermodynamically complex states frequently encountered in stellar astrophysics. We call this model the generalized pseudo-incompressible equations (GPI equations; 
\sec{Lagrangian Derivation of the Pseudo- Incompressible Equations}). More generally, we provide a systematic framework for producing approximate equations that filter fast dynamics.  In Appendix~\ref{sec: Constrained Magnetohydrodynamics} we apply this to the magnetohydrodynamic version of the GPI equations.  In the future, we expect to apply the Lagrangian framework to many other similar problems in astrophysical fluid dynamics such as rapidly rotating, strongly sheared, and/or multiple-component systems.  

Simple wave propagation problems provide a good test of the properties of different equation sets. In \sec{isothermal atmosphere} we derive and illustrated the behavior of gravity waves in the fully compressible (FC), pseudo-incompressible (PI/GPI), and low Mach number (LM) equations. We demonstrate that in an isothermal atmosphere, the energy flux in FC and PI/GPI gravity waves remains  constant with height, but that the LM gravity wave flux diverges with height. These properties
follow from energy conservation and non-conservation, respectively. 

In \sec{numerical} we conduct direct numerical simulations of linear gravity waves in isothermal atmospheres using two versions of the MAESTRO code; the standard version, which employs the LM equations, and a modified version which solves the  PI/GPI equations. We showed in Figures 4 and 5 that the eigenfunctions, energy conservation properties, and pseudo-energy conservation properties are as expected. Thus, we show that implementation of the PI/GPI equations is achievable and we hope that it will prove fruitful.

In this paper and in Part~I we worked only with ideal equations which omit viscosity, diffusion, and any other non-adiabatic heat transport effects. While it might be argued that energy conservation is only important in ideal systems, we note that if an ideal system does not conserve energy, there is no guarantee that non-ideal effects will decrease energy; they might cause spurious instabilities or other poorly controlled behavior. Thus, we recommend always working with equations which conserve energy in the ideal limit.

\acknowledgements

We thank Ann Almgren and John Bell, for extensive discussions about the pseudo-incompressible equations, low-Mach number equations and the MAESTRO code. We thank them additionally for permitting the use of MAESTRO in the demonstrations shown in \sec{bounded atmospheres and numerics}. 
We also thank Eliot Quataert, Nic Brummell, and Pascale Garaud for support and discussions, and additionally Keith Julien, Edgar Knobloch, and Mark Rast for helpful conversations.  
All authors found the well-known, but un-published, 2007 Mathematical Tripos, Part-III notes of Gordon Ogilvie invaluable in learning the salient points of Lagrangian Fluid Mechanics and MHD.
GMV acknowledges support from a TAC fellowship, and a CITA postdoctoral fellowship.
DL acknowledges support from a Hertz Foundation Fellowship, and the National Science Foundation Graduate Research Fellowship under Grant No. DGE 1106400.
TSW acknowledges support from NSF CAREER grant 0847477, and NSF grant NSF-0933759, and through the Center for Momentum Transport and
Flow Organization (CMTFO) sponsored by the US Department of Energy --- Office of Fusion Energy Sciences and the American Recovery and Reinvestment Act 2009. 
BPB is supported in part by NSF Astronomy and Astrophysics postdoctoral fellowship AST 09-02004. 
EGZ thanks the Department of Astronomy at U.~Chicago for their hospitality; a portion of this work was completed there.   The Center for Magnetic Self Organization (CMSO) is supported by NSF grant PHY 08-21899.  The simulations were carried out with NSF PACI support of NICS and TACC.

\appendix
\section{Radiation-Gas Equations of State}
\label{sec: radiative hydro}

Astrophysical fluids can depart significantly from the ideal gas equation of state assumption.  In this appendix we show how to implement a more complicated equation of state into the pressure-constrained GPI equations.  That is, we show how to compute $\Gamma_{1}(p_{0}(\vec{x}),\rho)$ in a non-trivial situation. 

As an example, we present a mixed radiation and gas model, which commonly applies in astrophysical systems ranging from the interiors of massive stars to accretion disks around black holes.  In this case, 
\Beq
p_{g} = \frac{k_{\mathrm{B}}}{m_{\mathrm{p}}} \rho T, \
p_{r} = (\gamma_{r}-1) a T^{\frac{\gamma_{r}}{\gamma_{r}-1}}, \
 p &=& p_{g} + p_{r} \quad \quad
\Eeq
specify the (ideal) gas, (blackbody) radiation, and combined total pressures respectively ($k_{\mathrm{B}}$, $m_{\mathrm{p}}$, and $a$ represent the Boltzmann constant, mass per gas particle, $a$ the radiation constant respectively).  For the radiation adiabatic exponent $\gamma_{r} = 1 + 1/\mbox{\textit{n}}$ where $\mbox{\textit{n}}$ is the dimensionality of the space (e.g., $\gamma_{r}=3/2$ in 2D and $\gamma_{r}=4/3$ in 3D).

Equations for the internal energy $e$ and entropy $s$ follow from \eq{2nd Law},
\Beq
&& \rho e = \frac{p_{r}}{\gamma_{r} -1} + \frac{p_{g}}{\gamma_{g} -1  },\n
&& s = \frac{\gamma_{r} a   T^{\inv{\gamma_{r}-1}} }{\rho} + \frac{k_{\mathrm{B}}}{m_{\mathrm{p}}} \log\frac{ T^ {\inv{\gamma_{g}-1}}  }{\rho}.
\Eeq
The temperature parameterizes the various thermodynamic variables, implicitly giving $p = p(\rho,s)$.  In this case the first adiabatic index mixes contributions from gas and radiation, \ie
\Beq
\label{eq:gamma1 rad hydro}
&&\nonumber \Gamma_{1} =\n\nonumber &&\frac{ \gamma_{g}(\gamma_{r}-1)^{2} p_{g}^{2}+\gamma_{r}(\gamma_{g}-1)p_{r}\left( (2\gamma_{r} -1) p_{g} + \gamma_{r}p_{r} ) \right)}{(p_{r}+p_{g})\left((\gamma_{r}-1)^{2}p_{g}+\gamma_{r}(\gamma_{g}-1)p_{r} \right)}\n &&\quad  \ = \frac{32- 24X - 3X^{2} }{24-21 X },
\Eeq
where $X \equiv p_{g}/(p_g + p_r)$, $\gamma_{g} = 5/3$, and $\gamma_{r}=4/3$ for the last expression in \eq{eq:gamma1 rad hydro}.  \Eq{eq:gamma1 rad hydro} for $\gamma_{r}=4/3$ matches the result given in \citet{Cox&Giuli_stellar_structure}, which also presents a wide range of astrophysically relevant equations of state.  When gas pressure dominates $\Gamma_{1} \approx \gamma_{g} = 5/3$ (for a monatomic gas in three dimensions). When radiation pressure dominates $\Gamma_{1} \approx \gamma_{r} = 1 + 1/\mbox{\textit{n}} = 4/3$ (in three dimensions).  Importantly, when neither gas nor radiation pressure dominate, then $\Gamma_{1}$ depends on both $\{T,\rho\}$, or implicitly on any other independent pairs of thermodynamic variables such as $\{p,\rho\}$, or $\{\rho,s\}$.

When applying the pressure constraint $p=p_{0}$, we must invert the implicit relationship 
\Beq
\label{p0-p-constraint} p_{0}(\vec{x}) = \frac{k_{\mathrm{B}}}{m_{\mathrm{p}}} \rho T + \frac{a T^{4}}{3}
\Eeq
by solving \eq{p0-p-constraint} for 
$T = T(p_{0}(\vec{x}),\rho)$.  The complete formula requires the roots of a 4th-order polynomial, but one may easily numerically solve an equation of the form $y = x + x^{4}$ for $x$ in terms of $y$.  Finding $T$ determines $p_{g} = p_{g}(p_{0}(\vec{x}),\rho)$ and $p_{r}=p_{r}(p_{0}(\vec{x}),\rho)$ individually.  Substituting into \eq{eq:gamma1 rad hydro} gives $\Gamma_{1}(p_{0}(\vec{x}), \rho)$.

\section{Free Energy} \label{sec: Free Energy}

In this appendix we derive the desired entropy function to produce a leading-order quadratic energy invariant.  To start, we define
\Beq
\nonumber F &=& \frac{\rho |\u |^{2}}{2} + \rho \left( e(\rho,s)  + \phi - f(s) \right) \n && = \frac{\rho |\u |^{2}}{2} + A(\rho,s) - p_{0}. 
\Eeq
where
\Beq
A(\rho,s) \equiv  p_{0} + \rho \left( e(\rho,s) - f(s) + \phi\right),
\Eeq
represents the available potential energy.  For notational convince, we define the perturbation variables 
\Beq
s_{1} \equiv s - s_{0}, \quad  \rho_{1} \equiv \rho - \rho_{0}.
\Eeq
Expanding the available potential energy to second order in the perturbations, 
\Beq
\label{2nd-order A expansion} \nn &&A(\rho,s) = A_{0} +  \rho_{1} \pd{A}{\rho}\bigg|_{0} + s_{1} \pd{A}{s}\bigg|_{0} +  \frac{\rho_{1}^{2}}{2} \pd{^{2}A}{\rho^{2}}\bigg|_{0}  +\n &&    \rho_{1} s_{1}  \pd{^{2}A}{\rho\partial{s}}\bigg|_{0} +  \frac{s_{1}^{2}}{2} \pd{^{2}A}{s^{2}}\bigg|_{0} + \O{\rho_{1}^{3},s_{1}^{3},s_{1} \rho_{1}^{2},\rho_{1}s_{1}^{2}},\quad \quad
\Eeq
where $|_{0}$ denotes evaluating at $\rho=\rho_0$ and $s=s_{0}$.
Requiring 
\Beq
\label{A-rho} &&\pd{A}{\rho}(\rho_{0},s_{0}) = e_{0} + \frac{p_{0}}{\rho_{0}} + \phi - f(s_{0})  = 0\n
\label{A-s} && \pd{A}{s}(\rho_{0},s_{0})  = \rho_{0}\left( T_{0} - f^{\prime}(s_{0})\right) = 0,
\Eeq 
ensures that the linear (non-sign-definite) terms in  \eq{2nd-order A expansion} vanish, and the quadratic terms produce the first non-trivial contributions to the free energy. \Eqs{A-rho}{A-s} imply the following functional relationships
\Beq
\label{?ground state?} f(s_{0}(\vec{x})) = h_{0}(\vec{x}) + \phi(\vec{x}), \ \  f^{\prime}(s_{0}(\vec{x})) = T_{0}(\vec{x}),\quad\quad
\Eeq
for background enthalpy, $h_{0}$, and temperature, $T_{0}$.  \Eq{?ground state?} only allows a solution if the two relations do not contradict each other.  Checking this, 
\Beq
\nonumber \dd{ \left (h_{0} + \phi - f(s_{0})\right)} & = & \dd{h_{0}} -\frac{\dd{p_{0}}}{\rho_{0}} - \dd{f(s_{0})}  \n &=&  ( T_{0} - f^{\prime}(s_{0}) ) \dd{s_{0}}  = 0,\quad \quad \quad
\Eeq
the enthalpy form of \eq{Enthalpy 2nd Law} implies compatibility between the two relations in \eq{?ground state?}.  

Assuming \eqs{A-rho}{A-s}, we compute the leading-order term in the available energy,
\Beq
A(\rho_{0},s_{0}) = 0.
\Eeq

To compute the quadratic terms in \eq{2nd-order A expansion}, we first define the total background pressure differential,
\Beq
\dd{p_{0}} =  - \rho_{0}\dd{\phi} = \pd{p}{\rho}(\rho_{0},s_{0})  \dd{\rho_{0}}   + \pd{p}{s}(\rho_{0},s_{0}) \dd{s_{0}}, \quad \quad
\Eeq
and differentiation with respect to the gravitational potential function, 
\Beq
\frac{d}{d\phi} \equiv - \frac{\g \dot \grad{}}{|\g |^{2}}.
\Eeq
Together,
\Beq
\pd{p}{s}(\rho_{0},s_{0}) = - \left(\frac{d s_{0}}{d \phi}\right)^{-1}\left( \rho_{0} + c_{0}^{2} \frac{d \rho}{d\phi}\right).
\Eeq

Therefore, 
\Beq
&&\pd{^{2}A}{\rho^{2}}(\rho_{0},s_{0}) = \frac{c_{0}^{2}}{\rho_{0}}\n
&& \pd{^{2}A}{\rho\partial{s}}(\rho_{0},s_{0}) = \inv{\rho_{0}}\pd{p}{s}(\rho_{0},s_{0})\n
&& \pd{^{2}A}{s^{2}}(\rho_{0},s_{0}) = \rho_{0}\left(f^{\prime \prime}(s_{0}) - \pd{T}{s}(\rho_{0},s_{0}) \right).
\Eeq

Using
\Beq
f^{\prime}(s_{0}) = T_{0} = \pd{e}{s}(\rho_{0},s_{0}),
\Eeq
and differentiating with respect to $\phi$, 
\Beq
\left(\!f^{\prime \prime}(s_{0}) - \pd{T}{s}(\rho_{0},s_{0})\! \right)\! \frac{ d s_{0} }{d \phi} = \inv{\rho_{0}^{2}} \frac{ d \rho_{0} }{d \phi} \pd{p}{s}(\rho_{0},s_{0}).\quad \quad
\Eeq
To linear order, density and entropy perturbations generate pressure perturbations via
\Beq
p_{1} = \rho_{1}\, c_{0}^{2}   + s_{1} \pd{p}{s}(\rho_{0},s_{0})  + \O{\rho_{1}^{2},s_{1}^{2},\rho_{1}s_{1}}.
\Eeq
Putting everything together, we obtain the available potential energy to quadratic order in terms of pressure and entropy perturbations,
\Beq
\label{quadratic A} A(\rho,s) = \frac{ p_{1}^{2}}{2 \rho_{0} c_{0}^{2}}  - \left(\rho_{0} + c_{0}^{2}\frac{d\rho_{0}}{d\phi} \right)\! \frac{\phi_{1}^{2}}{2} + \mathrm{h.o.t.}\quad \quad
\Eeq
where
\Beq
\phi_{1}  \equiv s_{1} \frac{d \phi}{d s_{0}}.
\Eeq
The background stratification determines the sign of the second coefficient in \eq{quadratic A}, and therefore determines the static stability to linear perturbations.

\section{Buoyancy Frequency and Entropy}
\label{sec: General Buoyancy Frequency}

Simply from the properties of differentials, the following two relations hold for any three thermodynamic quantities,
\Beq
\label{Useful Thermo} \thermod{a}{b}{c} \thermod{b}{a}{c} = 1, \quad \thermod{a}{b}{c} \thermod{b}{c}{a} \thermod{c}{a}{b} = -1.\quad
\Eeq
Computing the entropy differential
\Beq
\dd{s} = \thermod{s}{p}{\rho} \dd{p} + \thermod{s}{\rho}{p} \dd{\rho},
\Eeq
\eq{Useful Thermo} implies
\Beq
\nonumber -\inv{\rho}\thermod{\rho}{s}{p} \dd{s} &=& - \frac{p}{\rho} \thermod{\rho}{s}{p}\thermod{p}{\rho}{s} \frac{\dd{p}}{p}  - \frac{\dd{\rho}}{\rho}\n &=& \frac{\dd{p}}{\Gamma_{ 1}(p,\rho) p}  - \frac{\dd{\rho}}{\rho}
\Eeq
Carrying this further, and using commutativity of second partial derivatives (Maxwell relations)
\Beq
&&\nn -\inv{\rho} \thermod{\rho}{s}{p} = \inv{v} \thermod{v}{s}{p} = \inv{v}\pd{^{2}{h}}{s\partial{p}} = \inv{v}\thermod{T}{p}{s}  = \n &&\nn - \inv{v} \thermod{T}{s}{p}  \thermod{s}{p}{T} =  - \frac{T}{v c_{p}} \thermod{s}{p}{T} =  \n &&\frac{T}{v c_{p}}  \pd{^{2}{g}}{T\partial{p}} = \frac{T}{v c_{p}} \thermod{v}{T}{p} \equiv \frac{T \alpha_{T}}{c_{p}}. 
\Eeq
Therefore,
\Beq
\label{ds-dp-drho} T \alpha_{T} \frac{\dd{s}}{c_{p}} = \frac{\dd{p}}{\Gamma_{1}(p,\rho) p}  - \frac{\dd{\rho}}{\rho}
\Eeq
where
\Beq
\alpha_{T} \equiv \inv{\rho}\thermod{\rho}{T}{p}
\Eeq
defines the thermal expansion coefficient at constant pressure, and 
\Beq
c_{p} \equiv T\thermod{S}{T}{p} = \thermod{h}{T}{p}
\Eeq
defines the specific heat capacity at constant pressure.  The Brunt-V\"ais\"al\"a frequency in terms of the entropy gradient and gravity follows from \eq{ds-dp-drho}, 
\Beq
N^{2} = -T \alpha_{T} \frac{\g \dot\grad{s}}{c_{p}} = \g \dot \left(\frac{\grad{\rho}}{\rho} - \frac{\grad{p}}{\Gamma_{1} p} \right).  
\Eeq
For an ideal gas, $\alpha_{T} = 1/T$, $\Gamma_{1} = \gamma$.

\section{Constrained Magnetohydrodynamics}
\label{sec: Constrained Magnetohydrodynamics}

Many astrophysical fluids are magnetized plasmas, and can be modelled by coupling the Euler equations with electrodynamics to derive the magnetohydrodynamic (MHD) equations.  Assuming the fluid is a perfect conductor, the magnetic field remains frozen into the fluid, and evolves according to the induction equation
\Beq
\label{induction} \pd{\vec{B}}{t} = \curl{\left(\u \cross\vec{B}\right)}.
\Eeq
The magnetic field influences the velocity evolution through the Lorentz force, $\mu_0^{-1}\left(\curl\vec{B}\right)\cross\vec{B}$, which appears in the momentum equation.

Linear perturbations around a background flow and magnetic evolve according to
\Beq
\label{delta-B-eq} \pd{\delta \vec{B}}{t} = \curl{\left(\delta \u \cross  \vec{B} + \u \cross \delta \vec{B}\right)}.
\Eeq
Parameterizing perturbations using a displacement vector $\vxi$ defined in \eq{delta-u}, one can integrate \eq{delta-B-eq} such that
\Beq
\label{delta-B} \delta \vec{B} = \curl{\left(\vxi\cross \vec{B}\right)}.
\Eeq
The special case where $\vxi = \u \delta t$, and $\delta \vec{B} = (\partial_{t} \vec{B})\delta t$ confirms \eq{induction}.

As in \sec{Lagrangian Analysis}, we can derive the MHD momentum equation using Hamilton's principle of stationary action.  To correctly incorporate the Lorentz force, we must take the Lagrangian to depend on $\vec{B}$ in addition to $\u $, $\rho$, and $s$.  Defining
\Beq
\vec{j} \ \equiv \ \pd{\mathcal{L}}{\u},
\quad
\vec{H} \ \equiv \ -\pd{\mathcal{L}}{\vec{B}}, 
\Eeq
the general MHD Euler--Lagrange equation becomes
\Beq \label{eqn:Mmomentum}
&&\nn \pd{\vec{j}}{t} + \div{\left(\u \vec{j}\right)} + \grad{\u }\dot \vec{j} + \n&&  -\rho \grad{\left(\pd{\mathcal{L}}{\rho}\right)} +\pd{\mathcal{L}}{s}\grad{s} =  \left(\curl{\vec{H}}\, \right)\cross\!\vec{B}.
\Eeq
Using the Lagrangian \citep{Lundgren_1963},
\Beq
\mathcal{L}_{\mathrm{MHD}} \ = \ \rho\left( \frac{|\u |^{2}}{2} -  e(\rho,s) -  \phi \right) - \frac{|\vec{B}|^2}{2\mu_0},
\Eeq
one may straightforwardly check that the left-hand side of \eq{eqn:Mmomentum} reproduces the Euler momentum equation, while the right-hand side of \eq{eqn:Mmomentum} equals the Lorentz force\footnote{see un-published 2007 Mathematical Tripos, Part-III notes of Gordon Ogilvie for an excellent introduction to Hamiltonian MHD}.

Furthermore, for any Lagrangian depending on only $\u $, $\vec{B}$, $\rho$, and $s$, a conserved Hamiltonian, and generalized pressure tensor, follow respectively 
\Beq
\mathcal{H} & \equiv & \u \dot\vec{j} - \mathcal{L}, \n
\mathbf{P} & \equiv & \left(\mathcal{L} - \rho \pd{\mathcal{L}}{\rho} - \vec{B}\dot\pd{\mathcal{L}}{\vec{B}}\right)\mathbf{I} +\vec{B}\pd{\mathcal{L}}{\vec{B}},
\Eeq
and satisfy
\Beq
\partial_t\mathcal{H}+\div\left(\mathcal{H}\u + \mathbf{P}\vec{.}\u \right)=0.
\Eeq
Thus, the Lagrangian derivation maintains energy conservation.

To eliminate the fast magneto-acoustic modes from the MHD equations, we first define a base state satisfying magnetohydrostatic balance, 
\Beq
\grad{p_{0}} + \rho_{0} \grad{\phi} \ = \ \frac{\left(\curl{\vec{B}_{0}}\right)\cross \vec{B}_{0}}{\mu_{0}}.
\Eeq
In \sec{Lagrangian Derivation of the Pseudo- Incompressible Equations}, we argued that sound waves rapidly equilibrate the pressure within a fluid parcel to that of its surroundings.  From this physical argument, we derived the GPI equations by constraining the system so $p(\rho,s)=p_0(\vec{x})$.  In the MHD context, magneto-acoustic waves rapidly equilibrate a fluid parcel's {\it total} pressure to that of its surroundings, where the total pressure comprises the sum of the gas pressure, and the magnetic pressure,
\Beq
\Pi(\rho,s,\vec{B}) \equiv p(\rho,s) + \frac{|\vec{B}|^{2}}{2\mu_{0} }
\Eeq
Thus, we eliminate magneto-acoustic waves from the MHD equations by imposing the constraint 
\Beq
\C(\rho,s,\vec{B},\vec{x}) \equiv \Pi(\rho,s,\vec{B}) - \Pi_{0}(\vec{x}), 
\Eeq
and using the constrained Lagrangian.
\Beq
\label{constrained-MHD-L} \mathcal{L} &=& \mathcal{L}_{\mathrm{MHD}} - \lambda\,\C(\rho,s,\vec{B},\vec{x}),
\Eeq
The full set of equations associated follow from \eq{constrained-MHD-L},
\Beq
\label{MHD-rho}&& \frac{D\rho}{Dt} + \rho \div{\u } = 0,\n
&&\nn \rho \frac{D \u }{Dt} + (\rho-\rho_{0})\grad{\phi} =  - \grad{\left( \overline{\Gamma}\, \Pi_{0}\lambda\right)} + \lambda \grad{\Pi_{0}} + \quad \quad \quad  \n && \frac{\vec{B}\dot \grad{\left((1+\lambda)\vec{B}\right)}}{\mu_{0}} - \frac{\vec{B}_{0}\dot \grad{\vec{B}_{0}}}{\mu_{0}},\quad \quad \quad \n
&&\frac{D \vec{B}}{Dt} + \vec{B} \div{\u } = \vec{B}\dot\grad{\u },\n
&& \overline{\Gamma}\, \Pi_{0} \div{\u } + \u \dot\grad{\Pi_{0}} = \frac{\left(\vec{B}\dot \grad{\u }\right)\dot\vec{B}}{\mu_{0}},
\Eeq
where
\Beq
\label{MHD-Gamma} \overline{\Gamma} \ \equiv \ \frac{\Gamma_{1} + 2\,\frac{|\vec{B}|^{2}}{2\mu_{0}p}}{1 + \frac{|\vec{B}|^{2}}{2\mu_{0}p}} 
\Eeq
gives the average adiabatic exponent.  If gas pressure dominates over magnetic pressure ($2p\gg|\vec{B}|^2/\mu_0$), then $\overline{\Gamma}\approx\Gamma_1$, whereas if magnetic pressure dominates over gas pressure ($|\vec{B}|^2/\mu_0\gg 2p$), then $\overline{\Gamma}\approx 2$.  These equations conserve the same energy as the MHD equations, but have a modified pressure tensor,
\Beq
\mathbf{P} \ = \ \left(\Pi_{0} + \overline{\Gamma} \Pi_{0} \lambda \right)\mathbf{I} - (1+\lambda) \vec{B} \vec{B}.
\Eeq
Importantly, we show in upcoming work that \eqss{MHD-rho}{MHD-Gamma} correctly reproduce the dynamics of Alfv\'{e}n waves, as well as two- and three-dimensional magnetic buoyancy instability results \citep{Acheson_1979}.

Potential vorticity conservation no longer holds in the presence of magnetism.  Recall from \sec{Potential vorticity conservation} that PV conservation follows from the explicit invariance of the Lagrangian with respect ot displacements in the form of \eq{s, rho invariant}.  A major implication of PV conservation is that the two-dimensional pair $\{\rho,s\}$ act as reduced phase-space coordinates for the three-dimensional phase-space momenta $\rho\vec{u}$. Each reduction in phase-space dependence implies one conserved momenta.  

In the case of magnetism, \eq{delta-B} implies that the three-components of $\vec{B}$ also act as phase-space coordinates, bringing the apparent total to 5.  It seems that magnetic field confuses the accounting of phase-space dimension.  However,
\Beq
\div \vec{B} = 0,
\Eeq
implies that only two of the additional components of \eq{delta-B} vary independently.  However, it still seems that we have one extra coordinate for three-dimensional dynamics.  The final relation between the apparent four remaining coordinates comes from the conservation of potential magnetism 
\Beq
\label{Potential-Mag} \left[\Dt{}\right]\left(\frac{\grad s \dot \vec{B}}{\rho}\right) = 0.
\Eeq
\Eq{Potential-Mag} contains no significant dynamical information, and simply constrains an over-specification of phase space accompanying  \eq{induction}.  Now however, \eqs{rho-eq}{s-eq} along with \eq{Potential-Mag} imply fully six-dimensional phase-space dynamics with no conserved momenta in the form of \eq{GPV conservation}.

\section{Modifications to the Maestro code}\label{sec: modifications}

The numerical tests in \sec{numerical} required modifying the MAESTRO code to implement the PI \eqss{eq:new-momentum}{eq:new-density}.
The necessary modification follows from MAESTRO's implementation of the velocity divergence constraint in the LM equations
\citep[see][for more details]{Nonaka_et_al_2010, Nonaka_et_al_2012}.

The MAESTRO code assumes a velocity evolution of the form
\Beq
\label{u-evolution} \pd{\u }{t}  \ = \ - \inv{\rho}\grad{p^{\prime}} + \vec{a},
\Eeq
where the acceleration,
\Beq
\vec{a} \ = \ - \u \dot
  \grad{\u } + \frac{\left(\rho - \rho_{0}\right)}{\rho} \g.
\Eeq
\Eq{u-evolution} separates the accelerations due to pressure, and the accelerations due to inertia and gravity (and other sources in general).  The MAESTRO code satisfies the constraint
\Beq\label{eqn:constraint}
\div{\left(\beta_{0}\u \right)} \ = \ 0
\Eeq
using a divergence-cleaning scheme, evolving the velocity in two steps.   

To advance the velocity $\u ^n$ by $\Delta t$, MAESTRO first computes $\vec{a}^{n}$, from the solution at the $n$th time step, $\u^{(n)}$, and $\rho^{(n)}$.  With this, the code advances the flow according to 
\Beq
\label{eqn:evolveacceleration}
\tilde{\u}^{(n+1)} \ = \  \u^{(n)} + \vec{a} \Delta t.
\Eeq
The code then removes (cleans) the portion of $\tilde{\u}^{(n+1)}$ not satisfying \eq{eqn:constraint}.  This requires solving the elliptic equation, 
\Beq\label{eqn:LMelliptic}
\div{\left(\frac{\beta_{0}}{\rho}\grad{\chi}\right)} \ = \ \div{\left(\beta_{0}\tilde{\u}^{(n+1)}\right)}
\Eeq
for $\chi$.  The new flow solution,
\Beq\label{eqn:LMdivclean}
\u^{(n+1)} \ = \ \tilde{\u}^{(n+1)} - \frac{\grad{\chi}}{\rho},
\Eeq
satisfies \eq{eqn:constraint}.  We can then identify $\chi=p^{\prime}/\Delta t$.

The PI equations also employ the constraint \eq{eqn:constraint}, but use the velocity evolution equation
\Beq
 \pd{\u }{t}  \ = \ - \frac{\beta_0}{\rho}\grad{\left(\frac{p^{\prime}}{\beta_0}\right)} + \vec{a}.
\Eeq
The MAESTRO code easily implements this new velocity evolution equation.  First, we solve for $\tilde{\u}^{(n+1)}$ in the same way as  \eq{eqn:evolveacceleration}.  We then solve a modified equation for $\hat{\chi}$,
\Beq\label{eqn:PIelliptic}
\div{\left(\frac{\beta_{0}^2}{\rho}\grad{\hat{\chi}}\right)} \ = \ \div{\left(\beta_{0}\tilde{\u} ^{(n+1)}\right)},
\Eeq
and  update $\u^{(n+1)}$ via
\Beq\label{eqn:PIdivclean}
\u^{(n+1)} \ = \ \tilde{\u}^{(n+1)} - \frac{\beta_{0}\grad{\hat{\chi}}}{\rho}.
\Eeq
We now identify $\hat{\chi}= p^{\prime}/(\beta_{0}\Delta t)$.  To implement the PI equations instead of the LM equations, we replace \eqs{eqn:LMelliptic}{eqn:LMdivclean} with \eqs{eqn:PIelliptic}{eqn:PIdivclean}, respectively.

The distinction between the original MAESTRO implementation and our modifications amounts to  simply replacing  
\Beq
\grad{\chi} \to \beta_{0}\grad{\hat{\chi}}.  
\Eeq
In both cases, we merely subtract a vector from the initial evolution \eq{eqn:evolveacceleration}.  In both cases, the divergence of that vector equals the divergence of the output from \eq{eqn:evolveacceleration}.  
The difference between the \textit{vortical} components of these vectors provides the difference between the two methods.  
That is, even though 
\Beq
\div{\left(\frac{\beta_{0}}{\rho}\grad{\chi}\right)} \ = \ \div{\left(\frac{\beta_{0}^2}{\rho}\grad{\hat{\chi}}\right)},\\ \nonumber
\Eeq
it is also true that
\Beq
\curl{\left(\frac{\grad{\chi}}{\rho}\right)} \ \ne \ \curl{\left(\frac{\beta_{0} \grad{\hat{\chi}} }{\rho}\right)},
\Eeq
unless
\Beq
\grad{\beta_{0}}\times\grad{\hat{\chi}} \ = \ 0,
\Eeq
which is not satisfied in general. Therefore, we find that the PI and LM equations possesses non-trivial differences (see \sec{numerical}).

\end{document}